\documentclass[useAMS,usenatbib]{mn2e} 

\usepackage{graphicx} 
\usepackage{tabularx} 
\usepackage{times} 
  
\title[Remnant evolution after a carbon-oxygen white dwarf merger]{Remnant evolution 
after a carbon-oxygen white dwarf merger} 
\author[S.-C. Yoon,  Ph. Podsiadlowski \& S. Rosswog ]{S.-C. Yoon$^{1,2}$\thanks{E-mail: 
scyoon@science.uva.nl (SCY); podsi@astro.ox.ac.uk (PhP); s.rosswog@iu-bremen.de (SR)},   
Ph. Podsiadlowski$^{3}$ and S. Rosswog$^{4}$ \\ 
$^{1}$Astronomical Institute "Anton Pannekoek", University of Amsterdam, Kruislaan 403, 1098 SJ,  
Amsterdam, The Netherlands  \\ 
$^{2}$Department of Astronomy \& Astrophysics, University of California, Santa Cruz, CA95064, USA \\
$^{3}$Department of Astrophysics, University of Oxford, Keble Road, Oxford OX1 3RH, UK\\ 
$^{4}$School of Engineering and Science, Jacobs University Bremen\thanks{formerly International University Bremen}, Campus Ring1,  
Bremen 28759, Germany } 
\begin{document} 
 
\date{Accepted/ Received} 
 
\pagerange{\pageref{firstpage}--\pageref{lastpage}} \pubyear{2007} 
  
\maketitle 
  
\label{firstpage}  
  
\begin{abstract}  
We systematically explore the evolution of the merger of two 
carbon-oxygen (CO) white dwarfs. The dynamical evolution of a 
$0.9~\mathrm{M_\odot} + 0.6~\mathrm{M_\odot}$ CO white dwarf merger is 
followed by a three-dimensional SPH simulation. The calculation uses a 
state-of-the art equation of state that is coupled to an efficient 
nuclear reaction network that accurately approximates all stages from 
helium burning up to nuclear statistical equilibrium. We use an 
elaborate prescription in which artificial viscosity is essentially 
absent, unless a shock is detected, and a much larger number of SPH 
particles than earlier calculations. Based on this simulation, we 
suggest that the central region of the merger remnant can, once it has 
reached quasi-static equilibrium, be approximated as a differentially 
rotating CO star, which consists of a slowly rotating cold core and a 
rapidly rotating hot envelope surrounded by a centrifugally supported 
disc. We construct a model of the CO remnant that mimics the results 
of the SPH simulation using a one-dimensional hydrodynamic stellar 
evolution code and then follow its secular evolution, where we include 
the effects of rotation on the stellar structure and the transport of 
angular momentum.  The influence of the Keplerian disc is implicitly 
treated by considering mass accretion from the disc onto the hot 
envelope. The stellar evolution models indicate that the growth of the 
cold core is controlled by neutrino cooling at the interface between 
the core and the hot envelope, and that carbon ignition in the 
envelope can be avoided despite high effective accretion rates. This 
result suggests that the assumption of forced accretion of cold matter 
that was adopted in previous studies of the evolution of double CO 
white dwarf merger remnants may not be appropriate. Specifically we 
find that off-center carbon ignition, which would eventually lead to 
the collapse of the remnant to a neutron star, can be avoided if the 
following conditions are satisfied: (1) when the merger remnant 
reaches quasi-static equilibrium, the local maximum temperature at the 
interface between the core and the envelope must be lower than the 
critical limit for carbon-ignition. (2) Angular-momentum loss from the 
central merger remnant should not occur on a time scale shorter than 
the local neutrino cooling time scale at the interface. (3) The 
mass-accretion rate from the centrifugally supported disc must be 
sufficiently low ($\dot{M} \la 
5\times10^{-6}...10^{-5}~\mathrm{M_\odot~yr^{-1}}$).  Our results 
imply that at least some products of double CO white dwarfs merger may 
be considered good candidates for the progenitors of Type Ia 
supernovae. In this case, the characteristic time delay between the 
initial dynamical merger and the eventual explosion would be $\sim 
10^5\,$yr. 
\end{abstract} 
 
\begin{keywords} 
{Stars: evolution --  
 Stars: white dwarf -- 
 Stars: accretion -- 
 Supernovae: general -- 
 } 
\end{keywords} 
 
\section[]{Introduction}

The coalescence of two carbon-oxygen (CO) white dwarfs with a combined 
mass in excess of the Chandrasekhar limit has long been considered a 
promising path towards a Type Ia supernova (SN Ia; \citealt{Iben84}; 
\citealt{Webbink84}).  Indeed, in the last few years, a few massive 
double CO white dwarf systems have been found that have periods short 
enough for them to merge within a Hubble time 
(e.g. \citealt{Napiwotzki02,Napiwotzki04}).  This 
double-degenerate (DD) scenario can also easily explain the lack of 
hydrogen and helium lines in most SN Ia spectra and the occurrence of 
SNe Ia both in old and young star-forming systems 
(e.g. \citealt{Branch95}). 
 
Theoretically, the final fate of double CO white dwarf mergers has 
been much debated.  Previous studies assumed that the dynamical 
disruption of the Roche-lobe filling secondary should lead to the 
formation of a thick disc around the primary white dwarf 
(\citealt{Tutukov79}; \citealt{Mochkovitch89,Mochkovitch90}). 
Therefore, accretion of CO-rich matter from the thick disc onto the 
central cold white dwarf has been studied for investigating the 
evolution of such mergers by many authors (\citealt{Nomoto85}; 
\citealt{Saio85,Saio98,Saio04}; 
\citealt{Piersanti03a,Piersanti03b}).  As accretion rates from the thick 
disc should be close to the Eddington limit ($\dot{M} \approx 
10^{-5}~\mathrm{M_\odot yr^{-1}}$), most of those studies concluded 
that carbon ignition in the envelope of the accreting white dwarf is 
an inevitable consequence of such rapid accretion of CO-rich matter. 
Once carbon ignites off-center, the burning flame propagates inwards 
on a relatively short time scale ($\sim 5000$ yr), and the CO white 
dwarf is transformed into an ONeMg white dwarf 
(\citealt{Saio85,Saio98}).  When the mass of the ONeMg white dwarf 
approaches the Chandrasekhar limit, electron capture onto Ne and Mg 
is expected to lead to the gravitational collapse of the white dwarf 
to a neutron star (\citealt{Nomoto91}; see \citealt{Dessart06} and
\citealt{Kitaura06} for recent studies of such collapse).

\begin{figure} 
\center 
\resizebox{1.0\hsize}{!}{\includegraphics{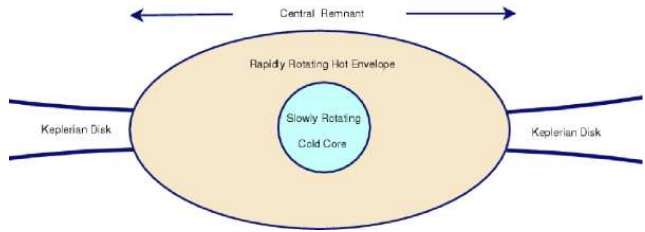}} 
\caption{Schematic illustration of the configuration of the remnant 
of a double CO white dwarf merger once quasi-static equilibrium 
has been established.} 
\label{figmerger} 
\end{figure}

However, the evolution of the remnants of double CO white dwarf 
mergers is not yet well understood.  For instance, it has been 
debated whether the accretion rate decreases when the accreting white 
dwarf reaches critical rotation \citep{Piersanti03a, Saio04}.  More 
importantly, the canonical description of the merger remnant as a 
primary white dwarf + thick disc system is clearly an 
oversimplification.  In previous three-dimensional smoothed particle 
hydrodynamics (SPH) simulations (\citealt{Benz90}; 
\citealt{Segretain97}; \citealt{Guerrero04}; see also 
Sect.~\ref{sectsph}), a large fraction of the disrupted secondary and 
the outermost layers of the primary form an extended hot envelope 
around the cold core containing most of the primary mass.  The rest of 
the secondary mass becomes a centrifugally supported disc in the 
outermost layers of the merger remnant.  Interestingly, the merger 
remnant reaches a state of quasi-static equilibrium within a few 
minutes from the onset of the dynamical disruption of the secondary. 
As the structure of the cold core plus the hot envelope appears to 
have a fairly spheroidal shape (see below) rather than the toroidal 
shape obtained with a zero-temperature equation of state 
\citep{Mochkovitch89,Mochkovitch90}, the merger remnant may be better 
described as a \emph{differentially rotating single CO star} 
consisting of a slowly rotating cold core and a rapidly rotating hot 
extended envelope surrounded by a Keplerian disc, as illustrated in 
Fig.~\ref{figmerger}, than the previously adopted primary white dwarf + 
thick disc system.  The further evolution of the merger must therefore 
be determined by the thermal cooling of the hot envelope and the 
redistribution of the angular momentum inside the central remnant, and 
accretion of matter onto the envelope from the Keplerian disc.

With this new approach to the problem in mind, we here revisit both 
the dynamical and the secular evolution of double CO white dwarf 
mergers. In the following section (Sect.~\ref{sectsph}), we present 
the numerical results of an SPH simulation of the dynamical evolution 
of the coalescence of a $0.9~\mathrm{M_\odot}$ WD and a 
$0.6~\mathrm{M_\odot}$ CO white dwarf up to the stage of 
quasi-hydrostatic equilibrium, and we carefully investigate the 
structure of the merger remnant. In Sect.~\ref{sectmerger}, we 
construct models of the central remnant in quasi-static equilibrium 
state (primary + hot extended envelope) which mimic the SPH result and 
calculate the thermal evolution of the merger remnant using a 
hydrodynamic stellar evolution code. In particular, the conditions for 
avoiding off-center carbon ignition are systematically explored. In 
Sect.~\ref{sectdiscussion}, we conclude this work by discussing 
uncertainties in our assumptions, the implications for Type Ia 
supernovae and future work.


\begin{figure*} 
\begin{center} 
\resizebox{0.40\hsize}{!}{\includegraphics{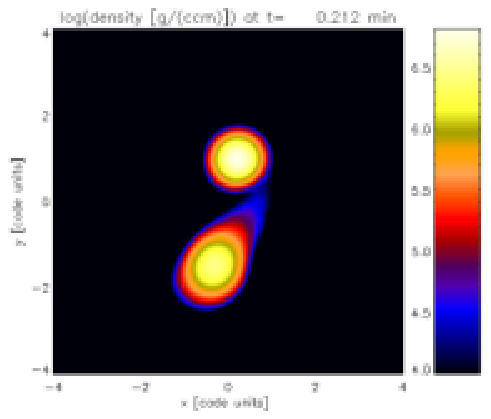}} 
\resizebox{0.40\hsize}{!}{\includegraphics{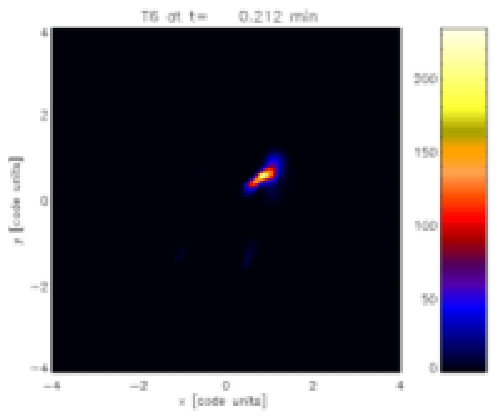}} 
\resizebox{0.40\hsize}{!}{\includegraphics{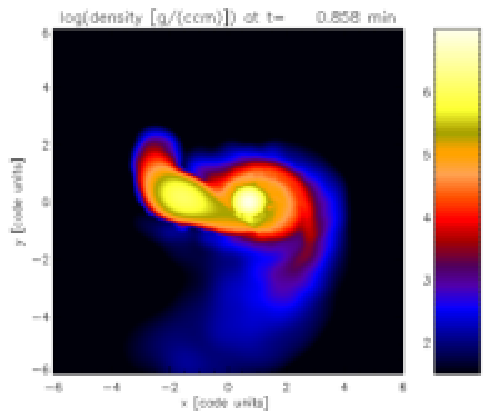}} 
\resizebox{0.40\hsize}{!}{\includegraphics{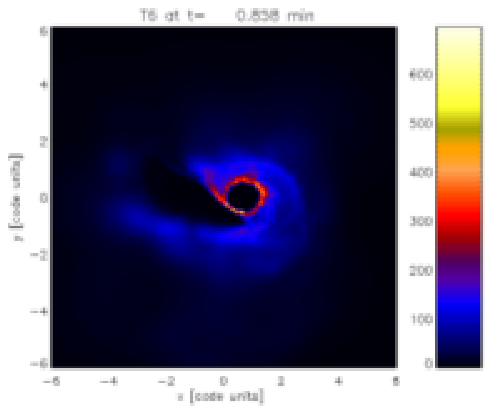}} 
\resizebox{0.40\hsize}{!}{\includegraphics{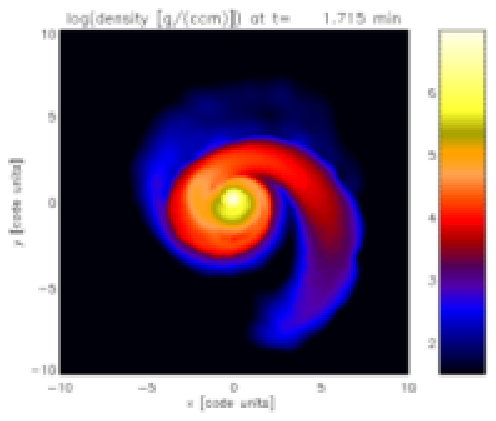}} 
\resizebox{0.40\hsize}{!}{\includegraphics{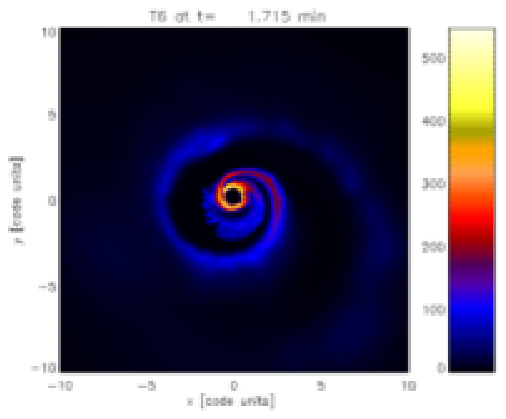}} 
\caption{Dynamical evolution of the coalescence of a  
$0.6~\mathrm{M_\odot} + 0.9~\mathrm{M_\odot}$ CO white dwarf binary. 
The panels in the left column show the density in the orbital plane, the panels 
in the right column the temperature in units of $10^6$~K.  
Lengths are in code units ($= 10^9~\mathrm{cm}$).  
}\label{figcoels1} 
\end{center} 
\end{figure*} 
 
\begin{figure*} 
\begin{center} 
\resizebox{0.40\hsize}{!}{\includegraphics{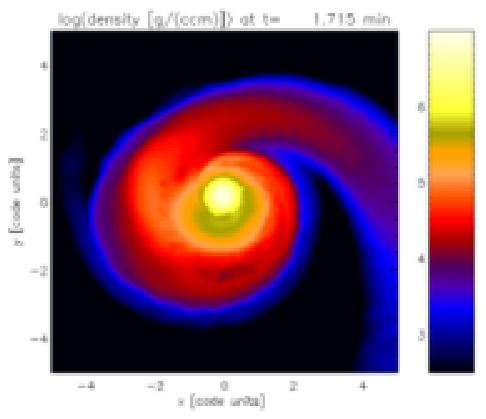}} 
\resizebox{0.40\hsize}{!}{\includegraphics{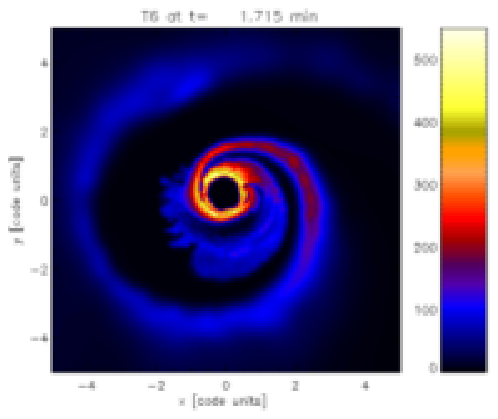}} 
\resizebox{0.40\hsize}{!}{\includegraphics{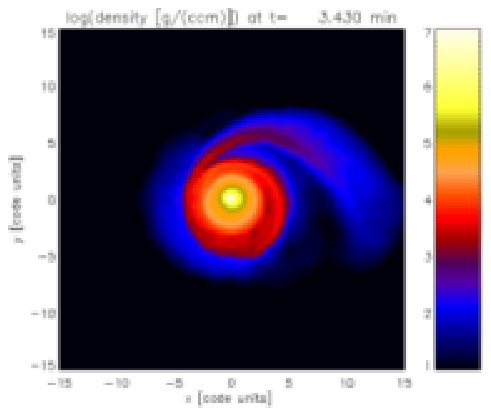}} 
\resizebox{0.40\hsize}{!}{\includegraphics{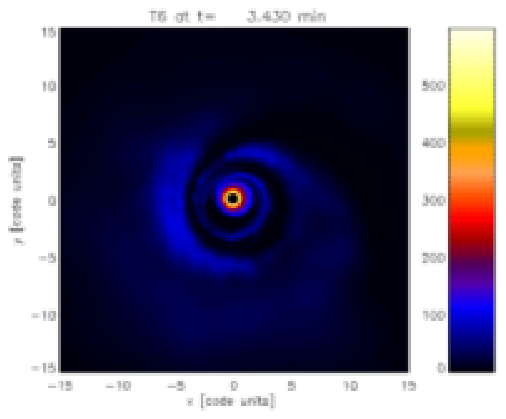}} 
\resizebox{0.40\hsize}{!}{\includegraphics{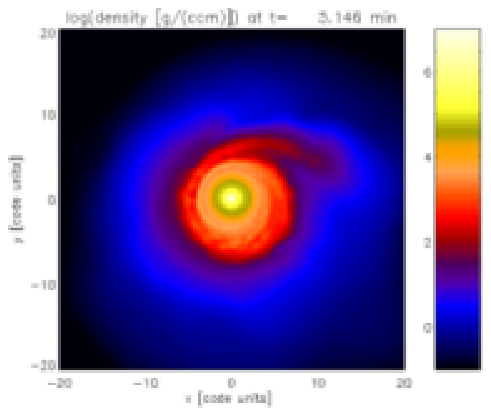}} 
\resizebox{0.40\hsize}{!}{\includegraphics{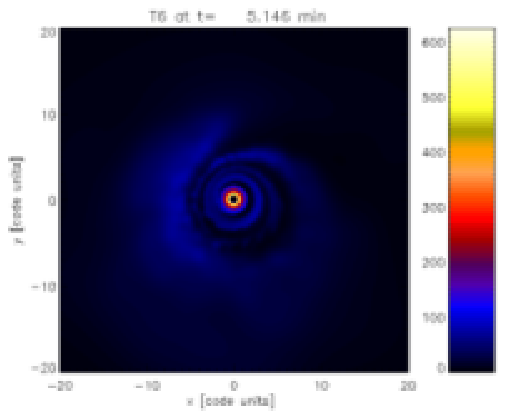}} 
\caption{Dynamical evolution of the coalescence of a  
$0.6~\mathrm{M_\odot} + 0.9~\mathrm{M_\odot}$ CO white dwarf binary.  
Continued from Fig.~\ref{figcoels1}.  
 }\label{figcoels2} 
\end{center} 
\end{figure*}

\section[]{Dynamical evolution of the merger}\label{sectsph}

Before discussing the subsequent thermal evolution after the 
coalescence of a double CO white dwarf coalescence binary, we 
investigate the configuration of the remnant in quasi-static 
equilibrium in some detail. For this purpose, we have carried out a 
SPH simulation of the dynamical process of the coalescence of two CO 
white dwarfs of $0.9~\mathrm{M_\odot}$ and $0.6~\mathrm{M_\odot}$, 
respectively. 
%
Our simulation uses a 3D smoothed particle hydrodynamics (SPH) code that 
is an offspring of a code developed to simulate neutron star mergers 
\citep{Rosswog00,Rosswog02,Rosswog03}. It uses an artificial  
viscosity scheme with time-dependent parameters \citep{Morris97}.  
In the absence of shocks, the viscosity parameters have a very low value  
($\alpha=0.05$ and $\beta= 0.1$; most SPH implementations use values of  
$\alpha= 1...1.5$ and $\beta= 2...3$); if a shock is detected, a source  
term \citep{Rosswog00} guarantees that the parameters rise 
to values that are able to resolve the shock properly without spurious  
post-shock oscillations. To suppress artificial viscosity forces in pure  
shear flows, we additionally apply a switch originally suggested by  
\citet{Balsara95}. 
 
To account for the energetic feedback onto the fluid from nuclear 
transmutations, we use a minimal nuclear reaction network developed by 
\citet{Hix98}.  It couples a conventional $\alpha$-network stretching 
from He to Si with a quasi-equilibrium-reduced 
$\alpha$-network. Although a set of only seven nuclear species is 
used, this network reproduces the energy generation of all burning stages from He-burning to 
NSE very accurately. For details and tests we refer to \citet{Hix98}. 
We use the HELMHOLTZ equation of state (EOS), developed by the Center 
for Astrophysical Thermonuclear Flashes at the University of 
Chicago. This EOS allows to freely specify the chemical composition of 
the gas and can be coupled to nuclear reaction networks. The 
electron/positron equation of state has been calculated without 
approximations, i.e. it makes no assumptions about the degree of 
degeneracy or relativity; the exact expressions are integrated 
numerically to machine precision. The nuclei in the gas are treated as 
a Maxwell-Boltzmann gas, the photons as blackbody radiation. The EOS 
is used in tabular form with densities ranging from $10^{-10} \le \rho 
Y_e \le 10^{11}~\mathrm{g~cm^{-3}}$ and temperatures from $10^4$ to 
$10^{11}$ K. A sophisticated, biquintic Hermite polynomial 
interpolation is used to enforce thermodynamic consistency (i.e. the 
Maxwell-relations) at interpolated values \citep{Timmes00}. 
 
We use a MacCormack predictor-corrector method 
(e.g. \citealt{Lomax01}) with individual particle time steps to evolve 
the fluid. With our standard parameters for the tree-opening criterion 
and the integration, this time marching implementation conserves the 
total energy to better than $4\times 10^{-3}$ and the total angular 
momentum to better than $2\times 10^{-4}$. Note that this could, in 
principle, be improved even further by taking into account the 
so-called ``grad-h''-terms \citep{Springel02,Monaghan02,Price04} and 
extra-terms arising from adapting gravitational smoothing terms 
\citep{Price06}. 
 
To avoid numerical artifacts, we only use equal mass SPH particles. For 
the initial conditions, we therefore stretch a uniform particle 
distribution according to a function that has been derived from 
solving the 1D stellar structure equations.  This technique is 
described in detail in \citet{Rosswog07}. This particle setup is then 
further relaxed with an additional damping force 
(e.g. \citealt{Rosswog04}) so that the particles can settle into their 
true equilibrium configuration. 
The calculations are performed with $2\times10^5$ SPH particles, a 
much larger particle number than could be afforded by previous 
calculations, and run up to a much longer evolutionary time (5 
minutes) than previous calculations (see Table~\ref{tabsph}).

\begin{figure} 
\begin{center} 
\resizebox{0.9\hsize}{!}{\includegraphics{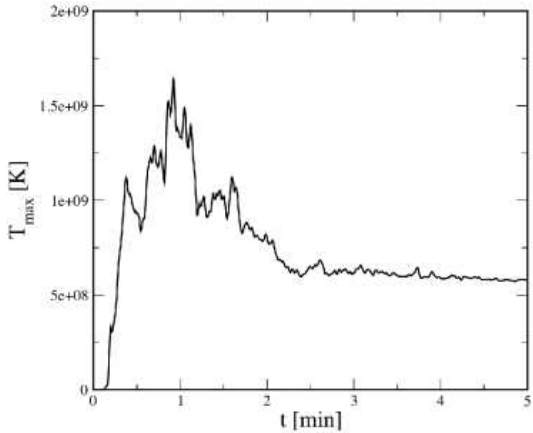}} 
\caption{The evolution of the local peak of temperature during the 
merger of two CO white dwarfs of $0.6~\mathrm{M_\odot}$ and 
$0.9~\mathrm{M_\odot}$, respectively, as a function of time after the 
onset of the simulation.  }\label{figtmax} 
\end{center} 
\end{figure}

\begin{figure} 
\begin{center} 
\resizebox{0.8\hsize}{!}{\includegraphics{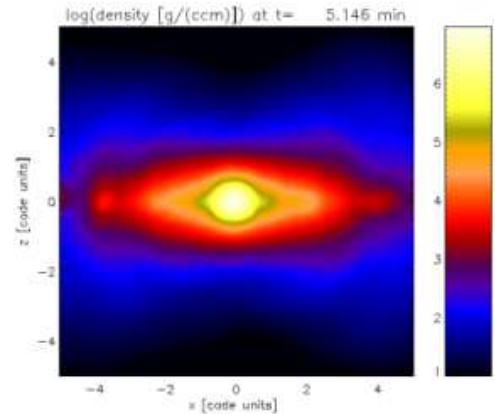}} 
\resizebox{0.8\hsize}{!}{\includegraphics{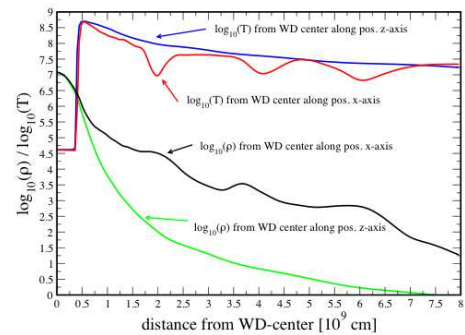}} 
\resizebox{0.8\hsize}{!}{\includegraphics{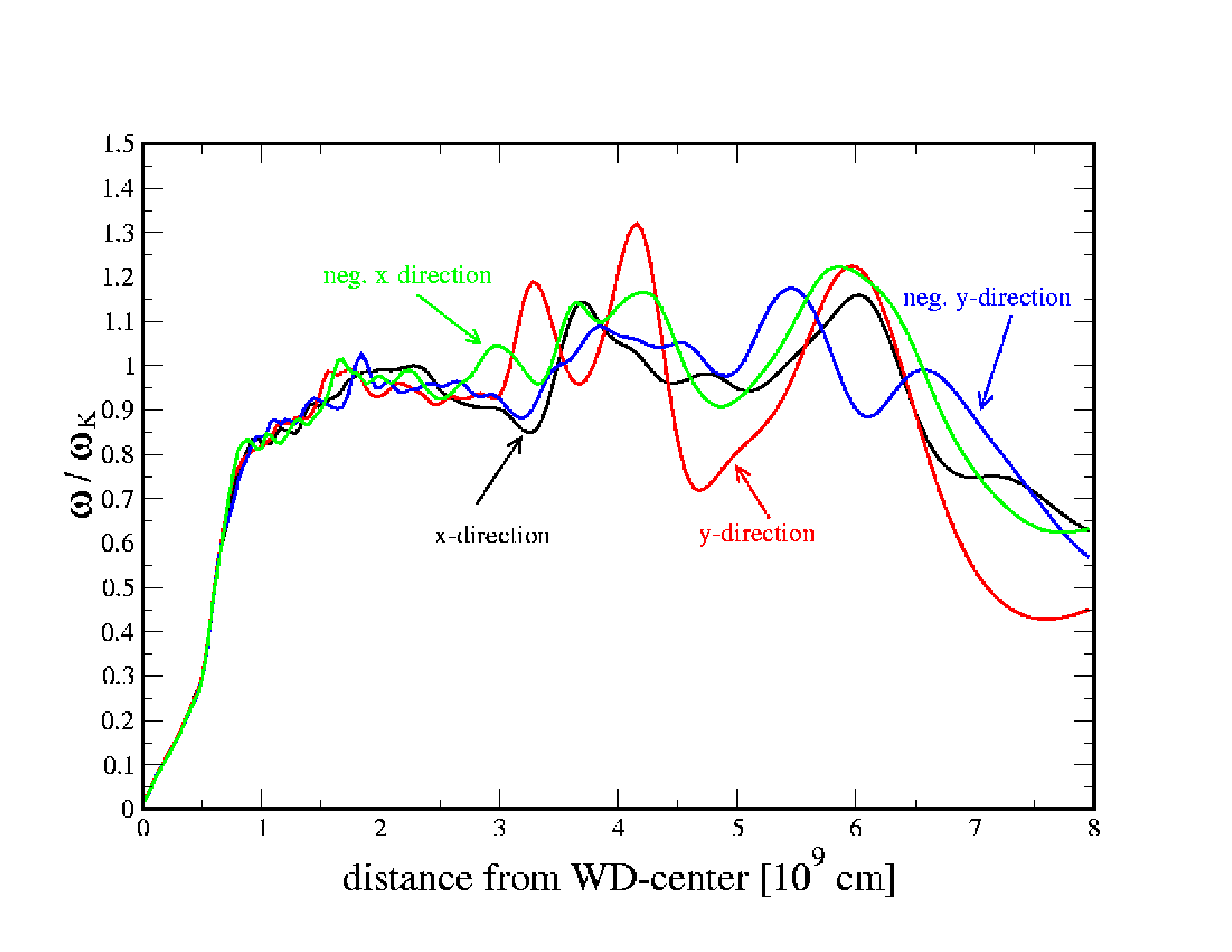}} 
\caption{ \emph{Top}: Density contour of the merger remnant in the 
$x-z$ plane at $t = 5.3~\mathrm{min}$. Here one code unit corresponds 
to $10^9~\mathrm{cm}$.  \emph{Middle}: Thermodynamic structure of the 
merger remnant at $t=5.31~\mathrm{min}$: shown are the temperature and 
the density as a function of distance from the centre, along the 
positive $x$- and $z$-axis, as indicated. \emph{Bottom}: Angular 
velocity in units of the local Keplerian value at 
$t=5.31~\mathrm{min}$, along the positive/negative $x$- and $y$-axis of 
the merger remnant.  }\label{figquasiequi} 
\end{center} 
\end{figure}

Figs.~\ref{figcoels1} and~\ref{figcoels2} show the dynamical evolution 
of the merging process of the double white dwarf system considered in 
this study. The panel in the left columns show the densities and the 
panel in the right columns the temperatures (in units of $10^6$ K) in 
the orbital plane.  The secondary is completely disrupted within 1.7 
minutes, and mass accretion onto the primary induces local heating 
near the surface of the primary.  Fig.~\ref{figtmax} shows the 
evolution of the maximum temperature as a function of time.  The peak 
in temperature reaches $1.7\times10^9~\mathrm{K}$ at $t\simeq 
1.0~\mathrm{min}$, where $t = 0.0$ marks the moment when the 
simulation starts. Carbon ignites when $T \ga 10^9~\mathrm{K}$, but 
nuclear burning is quenched soon due to the local expansion of the 
hottest region, as is also observed in the simulations of 
\citet{Guerrero04}.  The total amount of energy released due to 
nuclear burning is about $10^{45}~\mathrm{erg}$. 
 
\citet{Segretain97} considered the same initial white dwarf masses as 
in the present study. But they adopted the original artificial 
viscosity prescription of \citet{Monaghan88}, which is known to 
introduce spurious forces in shear flows, and they did not include 
nuclear burning (Table~\ref{tabsph}). By the end of their calculation 
($t=1.56~\mathrm{min}$), $T_\mathrm{max}$ reached 
$8\times10^8~\mathrm{K}$, while in our simulation, it decreases to 
$8\times10^8~\mathrm{K}$ only when $t\sim 1.7~\mathrm{min}$. 
Interestingly, $T_\mathrm{max}$ decreases further afterwards in our 
calculation, as shown in Fig.~\ref{figtmax}, and reaches a steady 
value at $T_\mathrm{max} \simeq 5.6\times10^8~\mathrm{K}$ when $t \ga 
\mathrm{2.5 min}$.  In the other calculations by \citet{Benz90} and 
\citet{Guerrero04}, the dynamical evolution of the merger was not 
followed for more than 2 minutes either, and we cannot directly 
compare our results to theirs.  However, we suspect that 
$T_\mathrm{max}$ would also decrease further in the systems they 
considered if they had continued their calculations for a longer 
evolutionary time.  It should also be noted that energy dissipation by 
artificial viscosity might lead to overheating, and that thermal 
diffusion -- which may play an important role in the outermost layers
-- is not considered in the present study.  It is thus likely that 
$T_\mathrm{max}$ in the quasi-static equilibrium state may be even lower in 
reality than in our simulation.

Fig.~\ref{figquasiequi} shows the structure of the merger remnant at 
quasi-static equilibrium.  The central region with $R \la 
10^9~\mathrm{cm}$ ($M_\mathrm{r}\la1.1~\mathrm{M_\odot}$) has a fairly 
spheroidal shape, and a centrifugally supported disc appears at $R \ga 
10^9~\mathrm{cm}$ where the angular velocity is close to the Keplerian 
value.  The fraction of the secondary mass contained in the Keplerian 
disc is larger in our simulation (about 67\%) than in 
\citet{Segretain97} (about 41 \%).  The innermost core ($R \la 3 
\times 10^8~\mathrm{cm}$; $M_\mathrm{r} \la 0.6~\mathrm{M_\odot}$) is 
essentially isothermal, and the temperature has its peak value 
($T\simeq5.6\times10^8~\mathrm{K}$) at $R\simeq 
5\times10^8~\mathrm{cm}$ and $M_\mathrm{r}\simeq 
0.85~\mathrm{M_\odot}$.  The disc material extends over 
$4\times10^9~\mathrm{cm}$ along the $z$-axis as the temperature is still 
high; if thermal diffusion were included, the disc would become much 
thinner on a short time scale of a few hours.  Therefore, our 
simulation confirms the remnant structure at quasi-static equilibrium 
that is illustrated in Fig.~\ref{figmerger}.  In the next section, we 
investigate the secular evolution of the merger from such a quasi-static 
equilibrium state.

\begin{table*} 
\begin{center} 
\caption{Comparison of SPH simulations of double CO white dwarf 
mergers. The columns list: $M_1$ and $M_2$: the 
masses of the primary and the secondary, respectively; NoP: the total 
number of particles used; $\nu_\mathrm{sph}$: the 
type of artificial viscosity employed, `std.' refers 
to \citet{Monaghan88}; Network: type of nuclear network employed; 
$t_\mathrm{sim}$: evolutionary time that has elapsed by  
the end of the calculation; $T_\mathrm{max}$: maximum temperature 
obtained during the simulation; and $T_\mathrm{p}$: the local peak of 
temperature at the end of the calculation}\label{tabsph} 
\vspace{0.1cm} 
\begin{tabularx}{\linewidth}{c c c l l c c c c} 
\hline 
\hline 
Ref.$^*$  & $M_1$                  & $M_2$                 & NoP & $\nu_\mathrm{sph}$  &  Network & $t_\mathrm{sim}$ & $T_\mathrm{max}$   & $T_\mathrm{P} $   \\ 
\hline 
 1 &$1.2~\mathrm{M_\odot}$&$0.9~\mathrm{M_\odot}$& $\sim 7\times10^3$& std. &  None & 51 sec. & ?  & $\sim 10^9~\mathrm{K}$  \\ 
 2 &$0.8~\mathrm{M_\odot}$&$0.6~\mathrm{M_\odot}$& $\sim 4\times10^4$& std. + Balsara-switch &  alpha network & 50 sec. &$1.4\times10^9~\mathrm{K}$ & ? \\ 
 2 &$1.0~\mathrm{M_\odot}$&$0.6~\mathrm{M_\odot}$& $\sim 4\times10^4$& std. + Balsara-switch &  alpha network & 65 sec. &$1.6\times10^9~\mathrm{K}$ & ? \\ 
 2 &$1.0~\mathrm{M_\odot}$&$0.8~\mathrm{M_\odot}$& $\sim 4\times10^4$& std. + Balsara-switch &  alpha network & 65 sec. &$2.0\times10^9~\mathrm{K}$ & ? \\ 
 3 &$0.9~\mathrm{M_\odot}$&$0.6~\mathrm{M_\odot}$& $\sim 6\times10^4$& std  &  None & 1.56 min. & ? &%
$\sim 7\times10^8~\mathrm{K}$  \\ 
 4 & $0.9~\mathrm{M_\odot}$ &$0.6~\mathrm{M_\odot}$ & $2\times10^5$ & see \cite{Rosswog00} &  QSE-alpha network & 5.3 min.  & $1.7\times10^9~\mathrm{K}$  & $5.6\times10^8~\mathrm{K}$  \\ 
\hline 
\end{tabularx} 
\end{center} 
$^{*}$1: \citet{Benz90}, 2: \citet{Guerrero04}, 3: \citet{Segretain97}; 4: Present Study 
\end{table*} 
 
\section[]{Secular evolution of the merger remnant}\label{sectmerger}

\begin{table*} 
\begin{center} 
\caption{Merger remnant model sequences. Each column lists the following: 
No.: sequence label, $M_\mathrm{CR}$: mass of the central 
remnant, $M_\mathrm{core}$: mass of the quasi-isothermal core, 
$M_\mathrm{p}$: location of the local peak of temperature in the mass 
coordinate, $T_{p}$: the local peak of temperature,
$\rho_\mathrm{p}$: density at $M_\mathrm{r} = 
M_\mathrm{p}$, $\tau_\mathrm{J}$: adopted time scale for angular 
momentum loss according to Eq.~(4), $\dot{M}_\mathrm{acc}$: adopted 
mass accretion rate from the Keplerian disc, C-ig: off-center ignition 
of carbon, $M_\mathrm{WD,ig}$: total mass of the central remnant when 
off-center carbon ignition occurs, $M_\mathrm{r, ig}$: location of 
off-center carbon ignition in the mass coordinate.}\label{tabmerger} 
\vspace{0.1cm} 
\begin{tabularx}{\linewidth}{c >{\centering\arraybackslash}X >{\centering\arraybackslash}X >{\centering\arraybackslash}X%
>{\centering\arraybackslash}X >{\centering\arraybackslash}X >{\centering\arraybackslash}X >{\centering\arraybackslash}X%
>{\centering\arraybackslash}X >{\centering\arraybackslash}X >{\centering\arraybackslash}X 
} 
\hline 
\hline 
No. & $M_\mathrm{CR}$ & $M_\mathrm{core}$ & $M_\mathrm{p}$ & $T_\mathrm{p}$ & $\rho_\mathrm{p}$ & $\tau_\mathrm{J}$ & $\dot{M}_\mathrm{acc}$ & C-ig & $M_\mathrm{WD,ig}$ & $M_\mathrm{r,ig}$ \\
  & [$\mathrm{M_\odot}$] & [$\mathrm{M_\odot}$]&  [$\mathrm{M_\odot}$]  & [$10^8~\mathrm{K}$] & [$10^6~\mathrm{g/cm^3}$] & [yr] & $\mathrm{10^{-6}~M_\odot/yr}$ &  & [$\mathrm{M_\odot}$] & [$\mathrm{M_\odot}$]  \\ 
\hline 
Sa1  &  1.11  & 0.6 & 0.84 & 5.6 & 0.8 &$\infty$ & 0.0  & No  & -   & -   \\ 
Sa2  &  1.11  & 0.6 & 0.84 & 5.6 & 0.8 &$10^2$ & 0.0  & Yes & 1.11& 0.80 \\ 
Sa3  &  1.11  & 0.6 & 0.84 & 5.6 & 0.8 &$10^3$ & 0.0  & Yes & 1.11& 0.80 \\ 
Sa4  &  1.11  & 0.6 & 0.84 & 5.6 & 0.8 &$10^4$ & 0.0  & Yes & 1.11& 0.85 \\ 
Sa5  &  1.11  & 0.6 & 0.84 & 5.6 & 0.8 &$10^5$ & 0.0  & No  & -   & -  \\ 
Sa6  &  1.11  & 0.6 & 0.84 & 5.6 & 0.8 &$10^5$ & 10.0 & Yes & 1.34 & 1.09  \\ 
Sa7  &  1.11  & 0.6 & 0.84 & 5.6 & 0.8 &$10^5$ &  5.0 & Yes & 1.34 & 1.20  \\ 
Sa8  &  1.11  & 0.6 & 0.84 & 5.6 & 0.8 &$10^5$ &  2.0 & No  &  -   &  -  \\ 
Sa9  &  1.11  & 0.6 & 0.84 & 5.6 & 0.8 &$10^5$ &  1.0 & No  &  -   &  -  \\ 
Sa10 &  1.11  & 0.6 & 0.84 & 5.6 & 0.8 &$5\cdot10^5$ &  5.0 & No  &  -   &  -  \\ 
Sa11 &  1.11  & 0.6 & 0.84 & 5.6 & 0.8 &$5\cdot10^5$ &  1.0 & No  &  -   &  -  \\ 
\hline 
Aa1  &  1.25  & 0.6 & 0.93 & 5.0 & 2.3 &$\infty$& 0.0 & No  &  -  &  - \\ 
Aa2  &  1.25  & 0.6 & 0.93 & 5.0 & 2.3 & $10^2$ & 0.0 & Yes & 1.250 & 0.90\\ 
Aa3  &  1.25  & 0.6 & 0.93 & 5.0 & 2.3 & $10^3$ & 0.0 & Yes & 1.250 & 0.92 \\ 
Aa4  &  1.25  & 0.6 & 0.93 & 5.0 & 2.3 & $10^4$ & 0.0 & Yes & 1.250 & 1.12 \\ 
Aa5  &  1.25  & 0.6 & 0.92 & 5.0 & 2.3 & $10^5$ & 0.0 & No  &  -    &  - \\ 
Aa6  &  1.25  & 0.6 & 0.92 & 5.0 & 2.3 & $10^6$ & 0.0 & No  &  -    &  -  \\ 
Aa7  &  1.25  & 0.6 & 0.92 & 5.0 & 2.3 & $10^5$ &10.0 & Yes & 1.360 & 1.20 \\ 
Aa8  &  1.25  & 0.6 & 0.92 & 5.0 & 2.3 & $10^5$ & 5.0 & No  &  -    &  - \\ 
Aa9  &  1.25  & 0.6 & 0.92 & 5.0 & 2.3 & $10^5$ & 1.0 & No  &  -    &  - \\ 
Aa10 &  1.25  & 0.6 & 0.92 & 5.0 & 2.3 & $10^6$ &10.0 & Yes & 1.382 & 1.22 \\ 
Ab1  &  1.25  & 0.7 & 0.92 & 5.0 & 3.1 &$\infty$& 0.0 & No  &   -   &  -  \\  
Ab2  &  1.25  & 0.7 & 0.92 & 5.0 & 3.1 & $10^3$ & 0.0 & Yes & 1.250 & 0.97 \\  
Ab3  &  1.25  & 0.7 & 0.92 & 5.0 & 3.1 & $10^4$ & 0.0 & No  &   -   &  -  \\  
Ab4  &  1.25  & 0.7 & 0.92 & 5.0 & 3.1 & $10^5$ & 0.0 & No  &   -   &  -  \\  
Ab5  &  1.25  & 0.7 & 0.92 & 5.0 & 3.1 & $10^5$ &10.0 & Yes & 1.344 & 1.21  \\  
Ab6  &  1.25  & 0.7 & 0.92 & 5.0 & 3.1 & $10^5$ & 5.0 & No  &   -   &  -  \\  
Ac1  &  1.25  & 0.5 & 0.88 & 5.9 & 1.6 &$\infty$& 0.0 & Yes & 1.250 & 0.84 \\ 
Ad1  &  1.25  & 0.6 & 0.92 & 6.0 & 1.8 &$\infty$& 0.0 & Yes & 1.250 & 0.90 \\ 
Ad2  &  1.25  & 0.6 & 0.92 & 6.0 & 1.8 & $10^6$ & 5.0 & Yes & 1.252 & 0.90 \\ 
Ae1  &  1.25  & 0.6 & 0.90 & 6.8 & 1.5 &$\infty$& 0.0 & Yes & 1.250 & 0.87\\  
\hline 
Ba1  & 1.363   & 0.82 & 0.95 & 5.0 & 12.2&$\infty$ & 0.0 & No  &  -  &  -  \\ 
Ba2  & 1.363   & 0.82 & 0.95 & 5.0 & 12.2&$10^2$ & 0.0 & Yes & 1.363 & 0.95  \\ 
Ba3  & 1.363   & 0.82 & 0.95 & 5.0 & 12.2&$10^3$ & 0.0 & Yes & 1.363 & 1.12  \\ 
Ba4  & 1.363   & 0.82 & 0.95 & 5.0 & 12.2&$10^4$ & 0.0 & No  &  -  &  -  \\ 
Ba5  & 1.363   & 0.82 & 0.95 & 5.0 & 12.2&$10^5$ & 0.0 & No  &  -  &  -  \\ 
Ba6  & 1.363   & 0.82 & 0.95 & 5.0 & 12.2&$10^5$ &10.0 & Yes & 1.398& 1.34  \\ 
Ba7  & 1.363   & 0.82 & 0.95 & 5.0 & 12.2&$10^5$ & 5.0 & Yes & 1.483& 1.43  \\ 
Ba8  & 1.363   & 0.82 & 0.95 & 5.0 & 12.2&$10^5$ & 1.0 & No  &  -  &  -  \\ 
\hline 
Ta1  & 1.25   & 0.60 & 0.86 & 5.0 & 28.8& $\infty$ & 0.0 & No & - & -\\ 
\hline 
\hline 
\end{tabularx} 
\end{center} 
\end{table*} 
 
\subsection[]{Physical assumptions and methods}\label{sectmethod} 
 
\begin{figure} 
\begin{center} 
\resizebox{0.8\hsize}{!}{\includegraphics{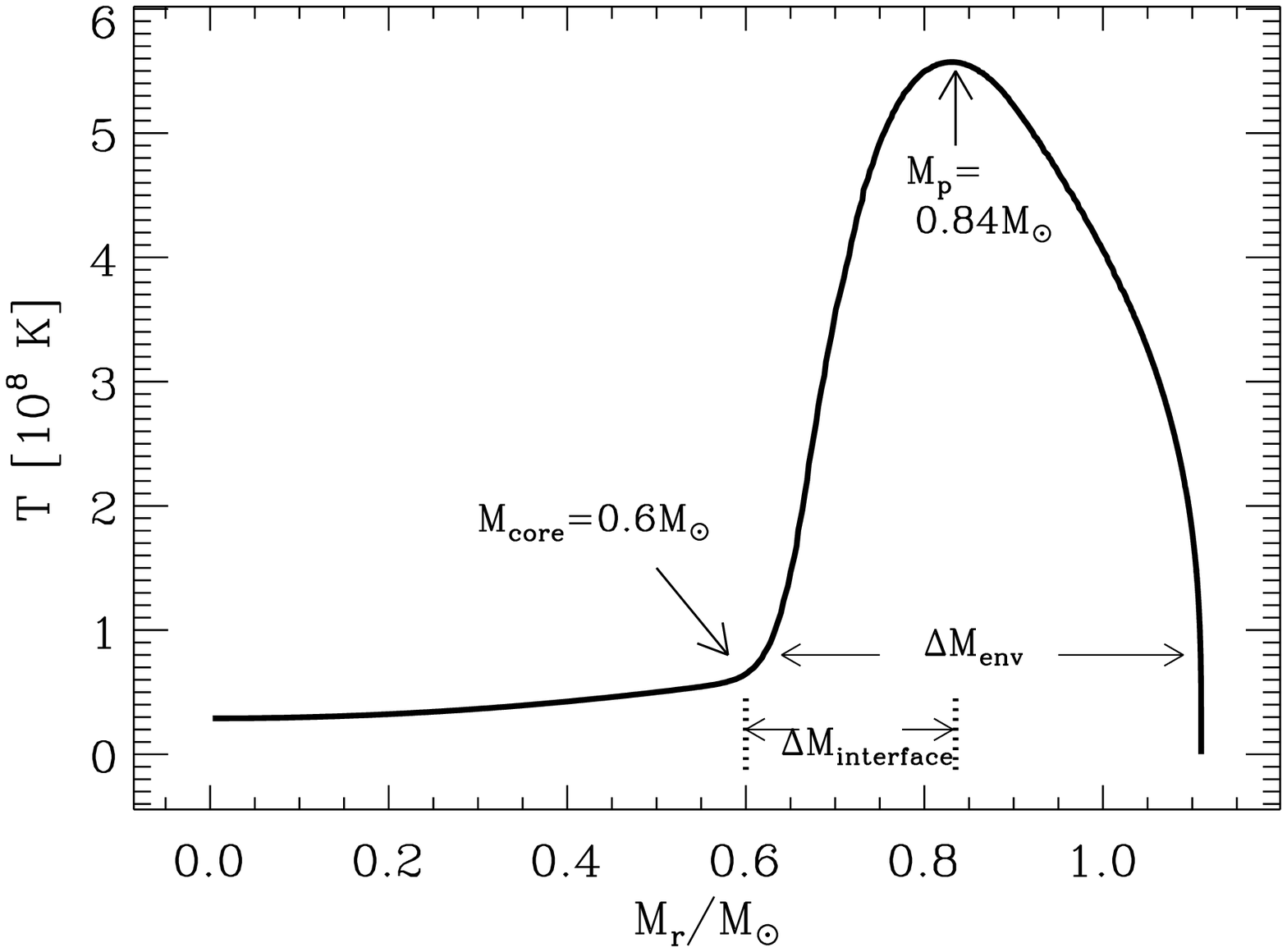}} 
\resizebox{0.8\hsize}{!}{\includegraphics{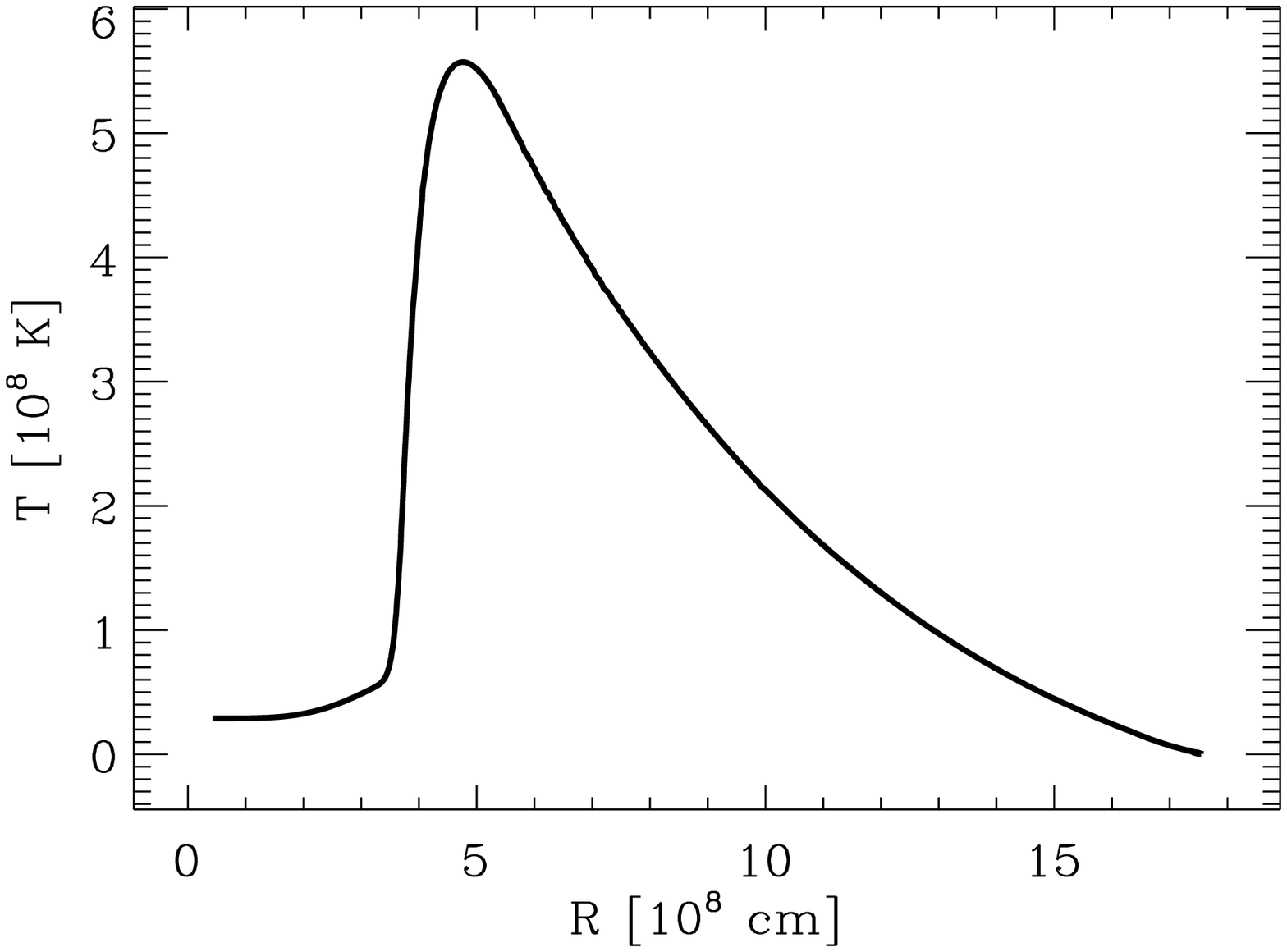}} 
\resizebox{0.8\hsize}{!}{\includegraphics{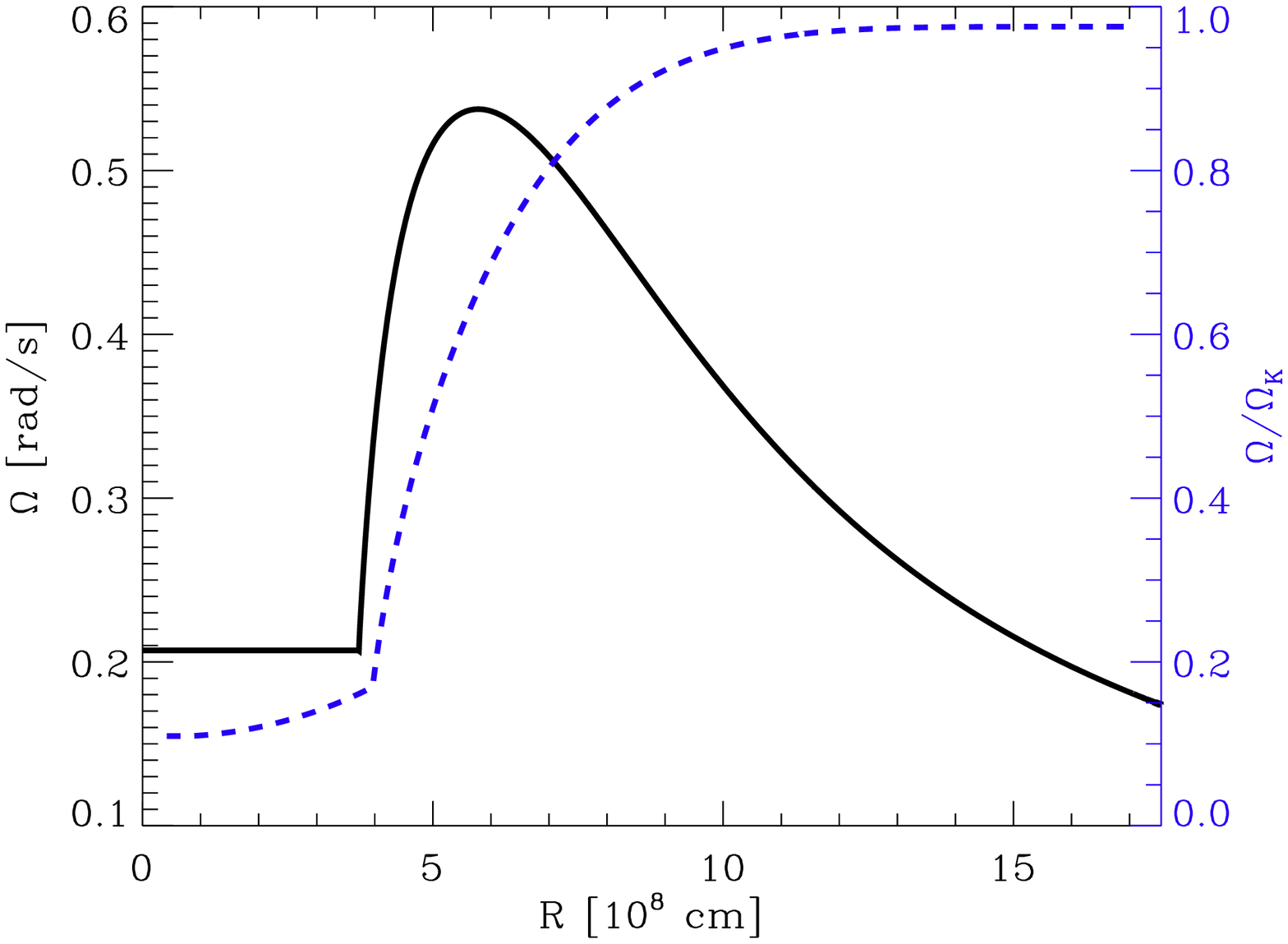}} 
\caption{Initial model of the central remnant for sequences Sa1 -- 
Sa11.  The top and middle panels show temperature as a function of the 
mass coordinate and radius, respectively.  The solid curve in the 
bottom panel gives the angular-velocity profile as a function of 
radius.  The dashed curve denotes the angular velocity in units of the 
local Keplerian value.  }\label{figinitial} 
\end{center} 
\end{figure}

Our SPH simulation shows that the remnant of the merger of two CO white 
dwarfs ($0.9~\mathrm{M_\odot} + 0.6~\mathrm{M_\odot}$) in the state of  
quasi-static equilibrium has the following features (see Fig.~\ref{figinitial}): 
\begin{enumerate} 
\item The core is cold and nearly isothermal. 
\item The local peak of temperature ($T_\mathrm{p}$) is located at a 
mass coordinate slightly less than the primary mass. 
\item A steep gradient in temperature appears at the interface between 
the core and the local peak of temperature. 
\item The interface is rather widely extended into the primary 
($\Delta M_\mathrm{interface} \approx$ 33 \% of the primary mass), and 
the mass of the quasi-isothermal cold core ($M_\mathrm{core}$) is 
about 77 \% of the primary mass. 
\item The mass of the outer envelope above the local peak of 
temperature contains about 33 \% of the mass of the secondary, and the 
rest of the secondary forms a Keplerian disc. 
\end{enumerate} 
 
Let us define $T_\mathrm{p}$ as the local peak of temperature at 
quasi-static equilibrium, $M_\mathrm{CM}$ as the mass of the central 
remnant (cold core + hot envelope), and $M_\mathrm{p}$ as the location 
of $T_\mathrm{p}$ in the mass coordinate (i.e., $M_\mathrm{p} = 
M_\mathrm{core}+\Delta M_\mathrm{interface}$; see 
Fig.~\ref{figinitial}).  To construct models of the central remnant, 
we use a one-dimensional hydrodynamic stellar evolution code which 
incorporates the effects of rotation on the stellar structure, 
transport of angular momentum due to the shear instability, 
Eddington-Sweet circulation, and the Goldreich, Schubert and Fricke 
instability, and dissipation of rotational energy due to shear 
motions. The effects of magnetic fields are neglected (see 
Sec.~\ref{sectdiscussion}). More details about the code are described 
in (\citealt{Yoon04}; hereafter YL04) and references therein.

\begin{figure*} 
\begin{center} 
\resizebox{0.4\hsize}{!}{\includegraphics{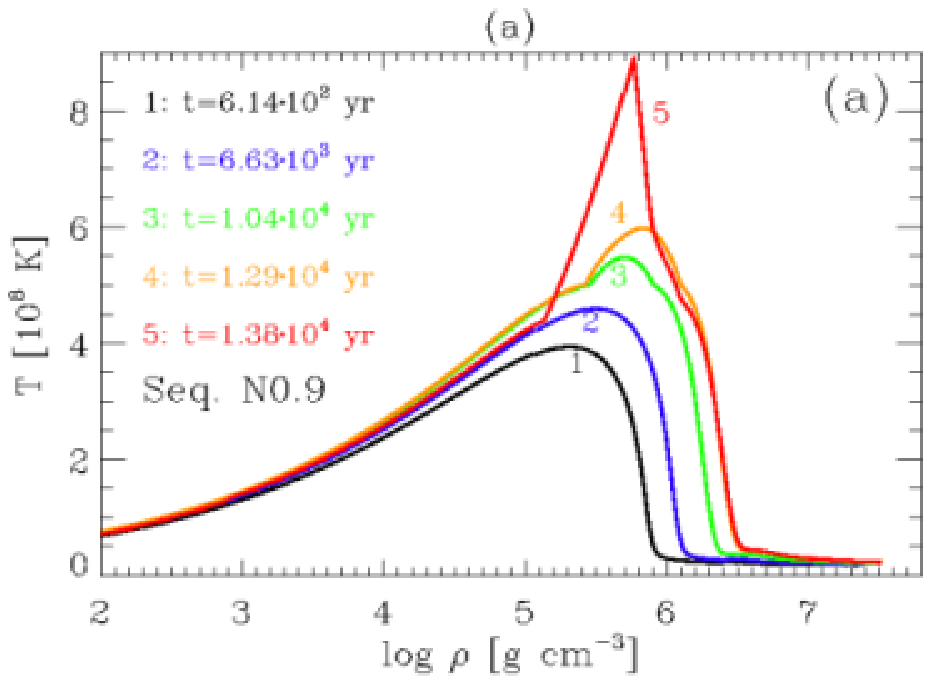}} 
\resizebox{0.4\hsize}{!}{\includegraphics{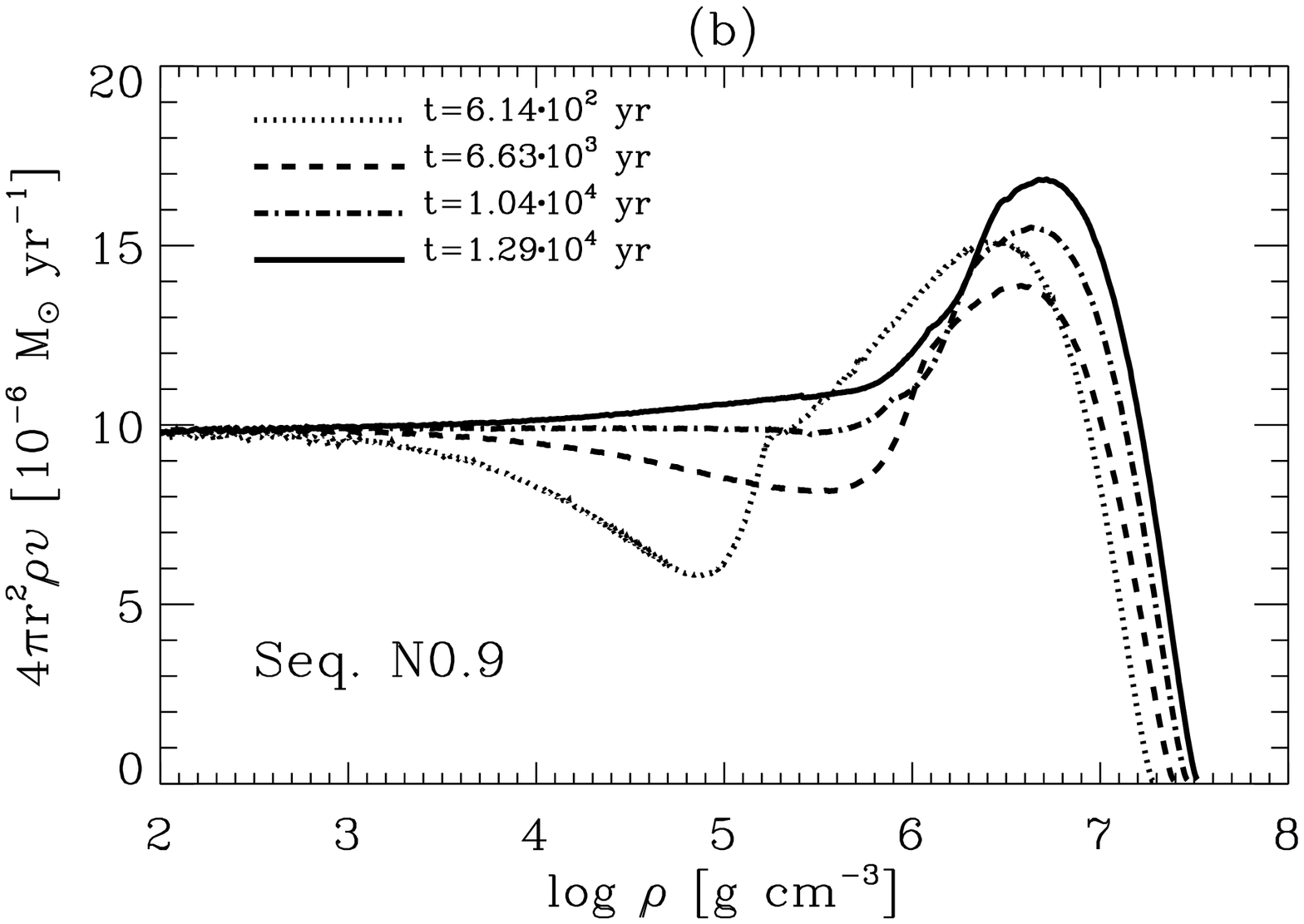}} 
\resizebox{0.4\hsize}{!}{\includegraphics{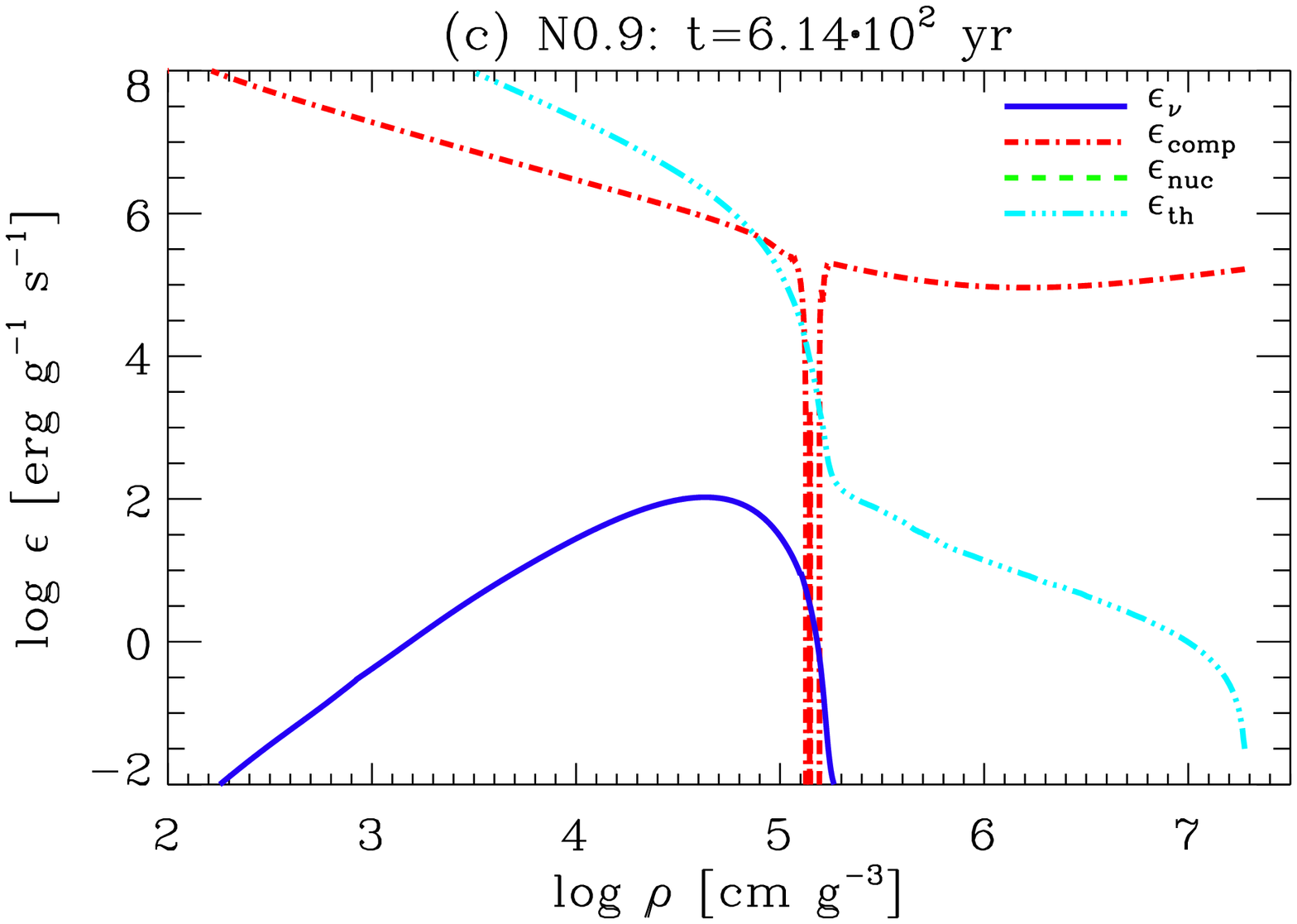}} 
\resizebox{0.4\hsize}{!}{\includegraphics{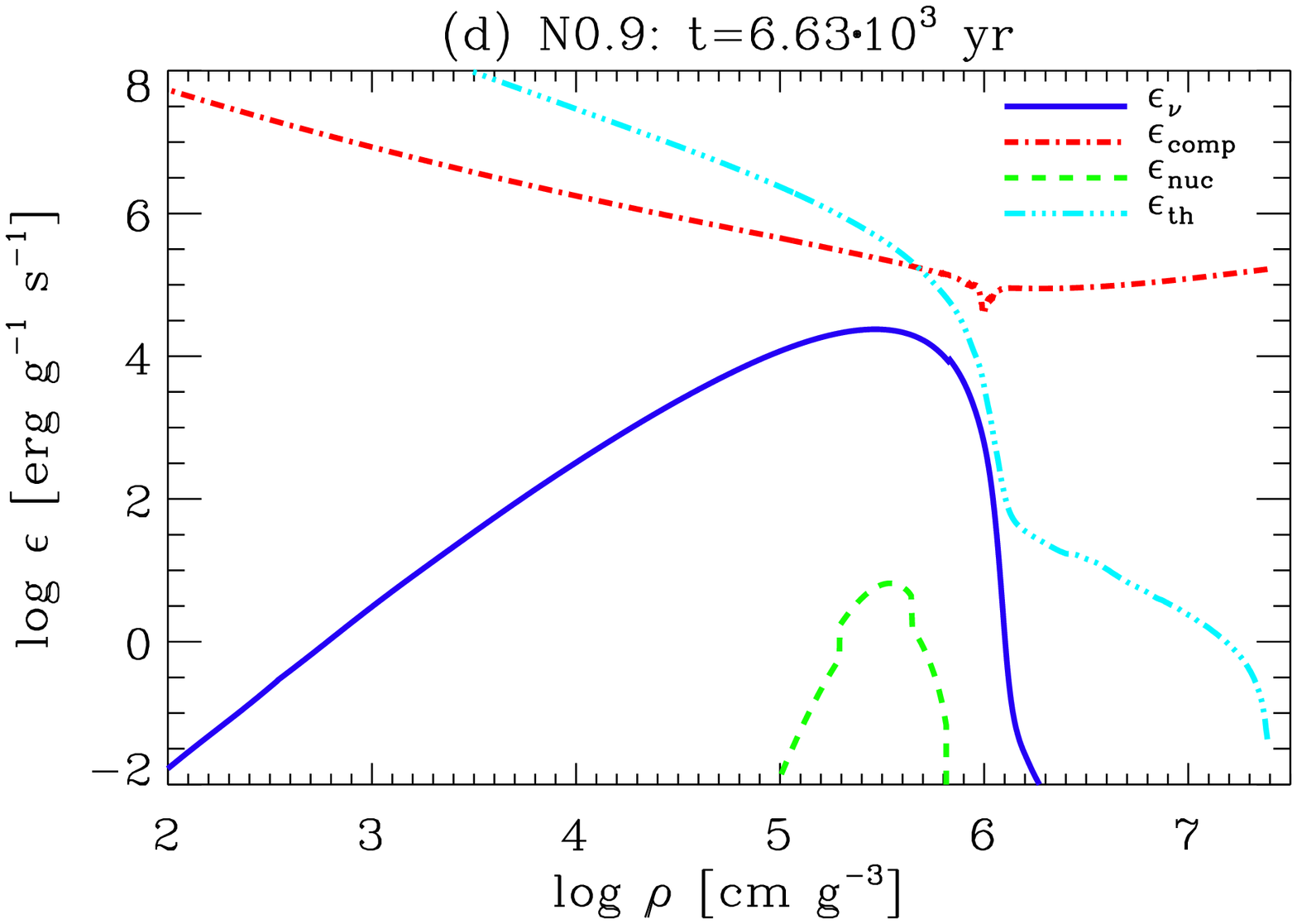}} 
\resizebox{0.4\hsize}{!}{\includegraphics{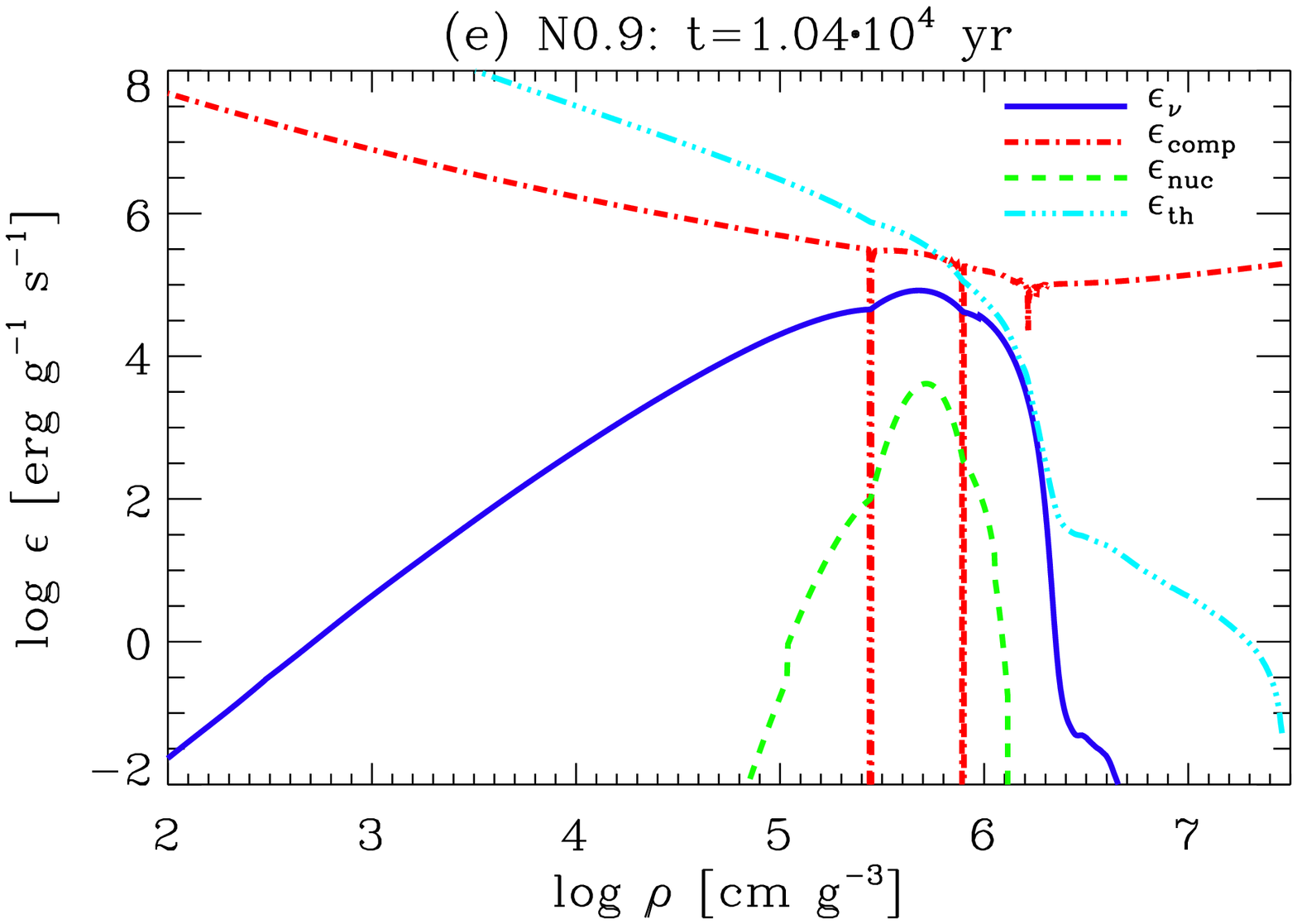}} 
\resizebox{0.4\hsize}{!}{\includegraphics{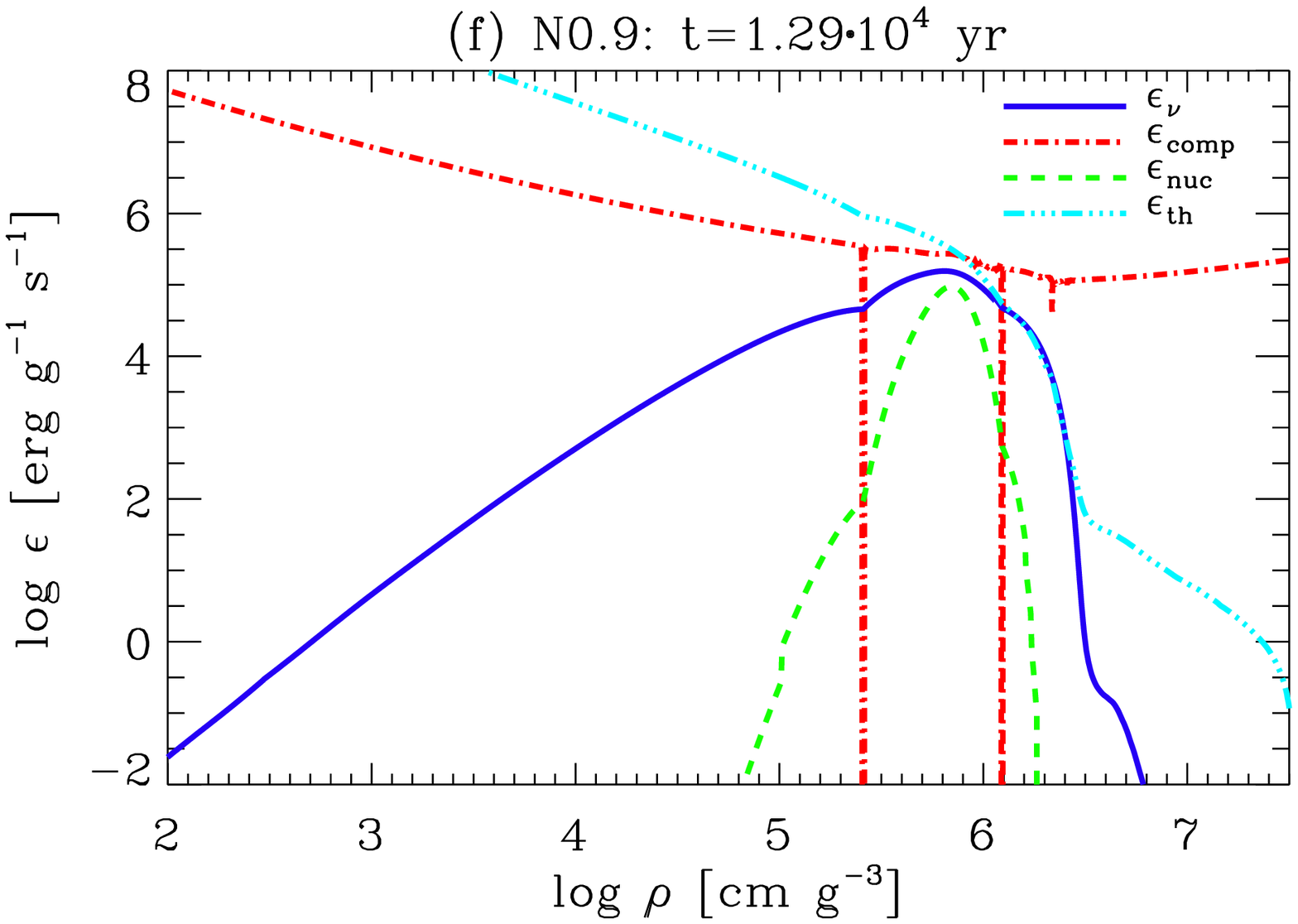}} 
\caption{(a) Evolution of a non-rotating white dwarf accreting with 
a constant accretion rate of $\dot{M} = 
10^{-5}~\mathrm{M_\odot~yr^{-1}}$ with an initial mass of 
$0.9~\mathrm{M_\odot}$ (Seq. N0.9) in the density -- temperature 
plane. (b) The local effective accretion rate ($\dot{M}_\mathrm{eff,r} := 
4\pi r^2\rho v$) as a function of density in Seq. N0.9, at different 
evolutionary epochs as indicated by the labels.  (c) -- (f) The rates of 
energy loss/production due to neutrino ($\epsilon_{\nu}$) cooling,  
compressional heating ($\epsilon_\mathrm{comp}$), nuclear energy generation 
($\epsilon_\mathrm{nuc}$) and thermal diffusion 
($\epsilon_\mathrm{th}$) at different evolutionary epochs. Note that 
here $\epsilon_{\nu}$, $\epsilon_\mathrm{comp}$ and 
$\epsilon_\mathrm{nuc}$ represent the values which are used in the 
evolutionary calculations, while $\epsilon_\mathrm{th}$ is an 
order-of-magnitude estimate according to Eq.~(7).  }\label{figN0.9} 
\end{center} 
\end{figure*}

In order to mimic the temperature profile of the central remnant as obtained 
from the SPH simulation, we artificially deposit energy in the envelope, 
using the following prescription for a white dwarf 
with $M=M_\mathrm{CR}$: 
\begin{equation} 
e(M_\mathrm{r}) = A(T'(M_\mathrm{r})-T(M_\mathrm{r}))~~[\mathrm{erg~g^{-1}~s^{-1}}], 
\end{equation} 
where 
\begin{equation} 
   \begin{array}{l} 
   T'(M_\mathrm{r})   =  \\ 
    
   \left\{ \begin{array}{ll}  
             3\cdot10^7\,{\rm K}+(7\cdot10^7\,{\rm K}-3\cdot10^7\,{\rm K} 
)\left(\frac{M_\mathrm{r}}{M_\mathrm{core}}\right)^2~, \\ 
\hspace{5cm} \textrm{if~$M_\mathrm{r}~< M_\mathrm{core}$}, \\ 
   T_\mathrm{p}-(T_\mathrm{p}-7\cdot10^7\,{\rm K})\left(\frac{M_\mathrm{p}-M_\mathrm{r}}%
   {M_\mathrm{p}-M_\mathrm{core}}\right)^2~,   \\
\hspace{4.5cm}        \textrm{if $M_\mathrm{core} \leq M_\mathrm{r} \leq M_\mathrm{p}$}, \\ 
   C - (C-T_\mathrm{p})\left(\frac{\log[\rho(M_\mathrm{r})/\rho_\mathrm{s}]}{\log [\rho(M_\mathrm{p})/\rho_\mathrm{s}]}\right)^2~, 
             \\ 
\hspace{5.5cm}              \textrm{if $M_\mathrm{r} > M_\mathrm{p}$}.  
           \end{array} \right. 
     \end{array}  
\end{equation} 
In this way, the temperature profile in the central  remnant model follows $T'(M_\mathrm{r})$.  
Here, $A$ and $C$ are constants. We use  
$A=10^5$\,erg\,g$^{-1}$\,s$^{-1}$\,K$^{-1}$  and $C =2\times10^8$\,K  
in most cases. 
 
A rotational profile is imposed as 
\begin{equation} 
   \begin{array}{l} 
  \Omega(M_\mathrm{r})  = \\ 
    \left\{\begin{array}{ll} 
     \Omega_\mathrm{O},  
\hspace{4.4cm}   \textrm{if $M_\mathrm{r} < M_\mathrm{core}$}, \\ 
     \left[\omega_\mathrm{O}+ (1-\omega_\mathrm{O})\left(\frac{M_\mathrm{r}-M_\mathrm{core}}%
         {M_\mathrm{CR}-M_\mathrm{core}}\right)^{0.9}\right]\sqrt{\frac{GM_\mathrm{r}}{r^3}},   \\
\hspace{5cm} \textrm{if $M_\mathrm{r} \geq M_\mathrm{core}$}, 
     \end{array} \right. 
   \end{array} 
\end{equation} 
where 
$\Omega_\mathrm{O} = 0.2\sqrt{GM_\mathrm{CR}/R^3}$ and
$\omega_\mathrm{O} = \Omega_\mathrm{O}/\sqrt{GM_\mathrm{core}/r_\mathrm{core}^3}$.
As shown in Fig.~\ref{figinitial}, this simple assumption gives a 
rotational velocity profile that is morphologically similar to that 
found in the SPH simulation: a steep gradient at the interface between 
the core and the envelope, and a local peak in the envelope.  Within 
our 1-D approximation of the effects of rotation, the exact shape of 
the rotational velocity profile does not affect the main conclusions of 
the present work for the following reasons. Firstly, the velocity 
gradient at the interface is adjusted to the threshold value for the 
dynamical shear instability on a very short time scale (see below, and 
discussions in YL04).  Secondly, our 1-D approximation underestimates 
the effect of the centrifugal force on the stellar structure in layers 
which rotate more rapidly than about 60 \% critical (YL04), and 
uncertainties due to this limit are much greater than due to the shape 
of $\Omega(r)$ in the outer layers of the envelope.  Possible 
uncertainties due to this limitation are critically discussed 
in Sect.~\ref{sectdiscussion}.

\begin{table} 
\begin{center} 
\caption{Accreting white dwarf model sequences with a constant 
accretion rate ($\dot{M}=10^{-5}~\mathrm{M_\odot/yr}$).  The columns 
list: No: sequence label, $M_\mathrm{init}$: 
initial mass, $\log L_\mathrm{init}/\mathrm{L}_{\odot}$ : initial 
luminosity, $M_\mathrm{WD, ig}$: the total mass of the white dwarf when carbon 
ignites off-center, and $M_\mathrm{r, ig}$: location of carbon ignition 
in the mass coordinate.  Sequences with `N' are for 
non-rotating models, and `R' for rotating models.  }\label{tabacc} 
\vspace{0.1cm} 
\begin{tabularx}{\linewidth}{c >{\centering\arraybackslash}X >{\centering\arraybackslash}X >{\centering\arraybackslash}X >{\centering\arraybackslash}X >{\centering\arraybackslash}X} 
\hline 
\hline 
No & $M_\mathrm{init}$ & $\log L_\mathrm{init}/\mathrm{L}_{\odot}$  & $M_\mathrm{WD, ig}$ & $M_\mathrm{r, ig}$  \\ 
\hline  
N0.7 & 0.7 & $-2.118$ & 0.999 & 0.793 \\ 
N0.8 & 0.8 & $-2.128$ & 1.010 & 0.862 \\ 
N0.9 & 0.9 & $-2.188$ & 1.039 & 0.939 \\ 
N1.0 & 1.0 & $-2.137$ & 1.087 & 1.024 \\ 
N1.1 & 1.1 & $-2.170$ & 1.150 & 1.114 \\ 
N1.2 & 1.2 & $-2.119$ & 1.225 & 1.207 \\ 
\hline 
R0.8 & 0.8 & $-2.114$ & 1.297 & 1.038 \\ 
R0.9 & 0.9 & $-2.119$ & 1.249 & 1.050 \\ 
R1.0 & 1.0 & $-2.082$ & 1.207 & 1.069 \\ 
R1.1 & 1.1 & $-2.050$ & 1.205 & 1.127 \\ 
\hline 
\end{tabularx} 
\end{center} 
\end{table}

The central remnant may lose angular momentum by outward angular 
momentum transport into the Keplerian disc 
\citep{Popham91,Paczynski91} and/or by the gravitational wave 
radiation, e.g., due to the $r$-mode instability 
\citep{Andersson98,Friedman98}.  Our code cannot properly describe any 
of these effects, and here we consider them simply by assuming a 
constant time scale for the angular momentum loss 
($\tau_\mathrm{J}$; see \citealt{Knaap04}; 
cf. \citealt{Piersanti03a}), such that the specific angular momentum 
of each mass shell decreases over a time step $\Delta t$ by an 
amount  
\begin{equation} 
\Delta j_i = j_i\left[1 - \exp(-\Delta t/\tau_\mathrm{J})\right]~. 
\end{equation} 
Mass accretion from the Keplerian disc is also considered in some 
model sequences, with different values for the accretion rate 
($\dot{M}_\mathrm{acc}$).  The angular-momentum accretion is treated 
in the same way as in YL04: the accreted matter is assumed to carry 
angular momentum at a value close to the Keplerian value if the 
surface velocity of the central remnant is below critical, while no 
angular-momentum accretion is allowed otherwise. 
 
Model sequences with different sets of $M_\mathrm{CR}$, 
$M_\mathrm{core}$, $M_\mathrm{p}$, $\tau_\mathrm{J}$, $\dot{M}$, and 
$T_\mathrm{p}$ are calculated, as summarized in Table~\ref{tabmerger}. 
The initial model in Seqs~S is intended to reproduce the result of 
our SPH simulation, where $M_\mathrm{CR} = 1.10~\mathrm{M_\odot}$ and 
$M_\mathrm{p}\approx0.84~\mathrm{M_\odot}$ are adopted.  We also 
assume $M_\mathrm{CR} = 1.25~\mathrm{M_\odot}$ and 
$M_\mathrm{p}\approx0.9~\mathrm{M_\odot}$ in Seq.~A, and 
$M_\mathrm{CR} = 1.364~\mathrm{M_\odot}$ and $M_\mathrm{p} = 
0.95~\mathrm{M_\odot}$ in Seq.~B, to simulate mergers of $0.9 - 1.0 
~\mathrm{M_\odot} + 0.7 - 1.0~\mathrm{M_\odot}$ white dwarf binaries. 
At a given $M_\mathrm{CR}$, different sets of $M_\mathrm{core}$, 
$M_\mathrm{p}$, and $T_\mathrm{p}$ are marked in the sequence label 
by minor characters (a, b, c, d, e), while different sets of 
$\tau_\mathrm{J}$ and $\dot{M}_\mathrm{acc}$ are indicated by Arabian 
numbers.  For instance, sequences Sa1 -- Sa11 have the same initial 
merger model, but different values for $\tau_\mathrm{J}$ and 
$\dot{M}_\mathrm{acc}$.  Rotation is neglected in a test sequence Ta1 
(i.e., the models are non-rotating).  The temperature and angular-velocity 
profiles in the initial central remnant model of Seqs Sa1 - Sa11 are 
shown in Fig.~\ref{figinitial}. The temperature (a few to several 
$10^8~\mathrm{K}$) and the size ($\sim 10^9~\mathrm{cm}$) of the envelope 
appear to be comparable to those obtained from the SPH simulation (see 
Fig.~\ref{figquasiequi}). 
 
For comparison, we also ran model sequences for classical cold-matter 
accretion with a constant accretion rate of $\dot{M} = 
10^{-5}~\mathrm{M_\odot/yr}$, for both non-rotating and rotating 
cases, as summarized in Table~\ref{tabacc}.

\subsection[]{Results}\label{sectresults} 
 
\subsubsection[]{Classical models of cold-matter accretion} 
 
Before discussing the central remnant models, let us first investigate 
the evolution of classical cold-matter accreting white dwarf models in 
detail.  In these models, the accreted matter is assumed to have the 
same entropy as the surface value of the accreting white dwarf.  As 
shown in previous studies (e.g. \citealt{Nomoto85}), the thermal evolution of rapidly accreting 
white dwarfs is determined by the interplay of compressional heating 
and thermal diffusion.  Fig.~\ref{figN0.9} shows an example of the 
evolution of such accreting white dwarf models for an initial WD mass 
of $0.9~\mathrm{M_\odot}$ and a constant accretion rate of 
$\dot{M}_\mathrm{acc} = 10^{-5}~\mathrm{M_\odot~yr^{-1}}$ (Seq. N0.9; 
Table~\ref{tabacc}). 
 
\begin{figure} 
\begin{center} 
\resizebox{0.8\hsize}{!}{\includegraphics{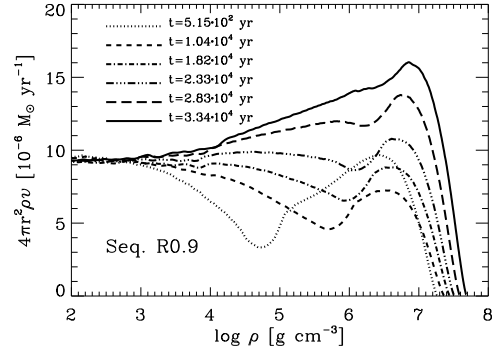}} 
\caption{Local effective mass accretion rate ($\dot{M}_\mathrm{eff, r} 
\equiv 4\pi r^2\rho v$) as a function of density in the models of sequence 
R0.9, at different evolutionary epochs.}\label{figmdotR0.9} 
\end{center} 
\end{figure}

Fig.~\ref{figN0.9}a shows that the temperature increases continuously 
in the envelope ($\rho \la 10^6~\mathrm{g~cm^{-3}}$), and finally 
carbon burning becomes significant at $\rho \simeq 
5.6~\times10^5~\mathrm{g~cm^{-3}}$ and $T \simeq 
6\times10^8~\mathrm{K}$ when $t \simeq 1.3\times10^4~\mathrm{yr}$.  In 
Figs.~\ref{figN0.9}c --f, the rates of compressional heating 
($\epsilon_\mathrm{comp}$), neutrino cooling ($\epsilon_{\nu}$), 
nuclear energy generation ($\epsilon_\mathrm{nuc}$) and thermal 
diffusion ($\epsilon_\mathrm{th}$) are shown.  In our stellar 
evolution code, the compressional heating rate is calculated according to 
\begin{equation} 
\epsilon_\mathrm{comp}=\frac{P}{\rho^2}\left(\frac{\partial\rho}{\partial t}\right)_{M_\mathrm{r}} %
\end{equation} 
\citep{Kippenhahn90}.  Neutrino cooling rates are obtained following 
\citet{Itoh96}.  While $\epsilon_\mathrm{comp}$, $\epsilon_{\nu}$, and 
$\epsilon_\mathrm{nuc}$ in the figures correspond to the values that 
are used for the evolutionary calculations, the thermal diffusion rate 
($\epsilon_\mathrm{th}$) -- which is only calculated implicitly in the 
code -- can only be estimated to within an order-of-magnitude from 
\begin{equation} 
\epsilon_\mathrm{th} \approx TC_\mathrm{P}/\tau_\mathrm{th}~~. 
\end{equation} 
Here $C_\mathrm{P}$ denotes the specific heat at constant pressure, and $\tau_\mathrm{th}$ the local 
thermal diffusion time scale defined as 
\begin{equation} 
\tau_\mathrm{th}\equiv H_\mathrm{P}^2/K~~, 
\end{equation} 
where $H_\mathrm{P}$ is the pressure scale height, and $K$ 
[$(4acT^3)/(3C_\mathrm{P}\kappa\rho^2)$] is the thermal diffusivity. 
It is clear from Fig.~\ref{figN0.9} that the local peak of temperature 
is located where the compressional heating rate begins to dominate 
over the thermal diffusion rate ($\rho \approx 
10^5~\mathrm{g~cm^{-3}}$), as expected.  The neutrino cooling rate 
also increases as the temperature in the envelope becomes higher, but 
nuclear energy generation becomes significant before neutrino cooling 
dominates the thermal evolution, inducing a carbon-burning flash 
around $\rho \simeq 5.6\cdot10^5~\mathrm{g~cm^{-3}}$. 
 
\begin{figure*} 
\begin{center} 
\resizebox{0.4\hsize}{!}{\includegraphics{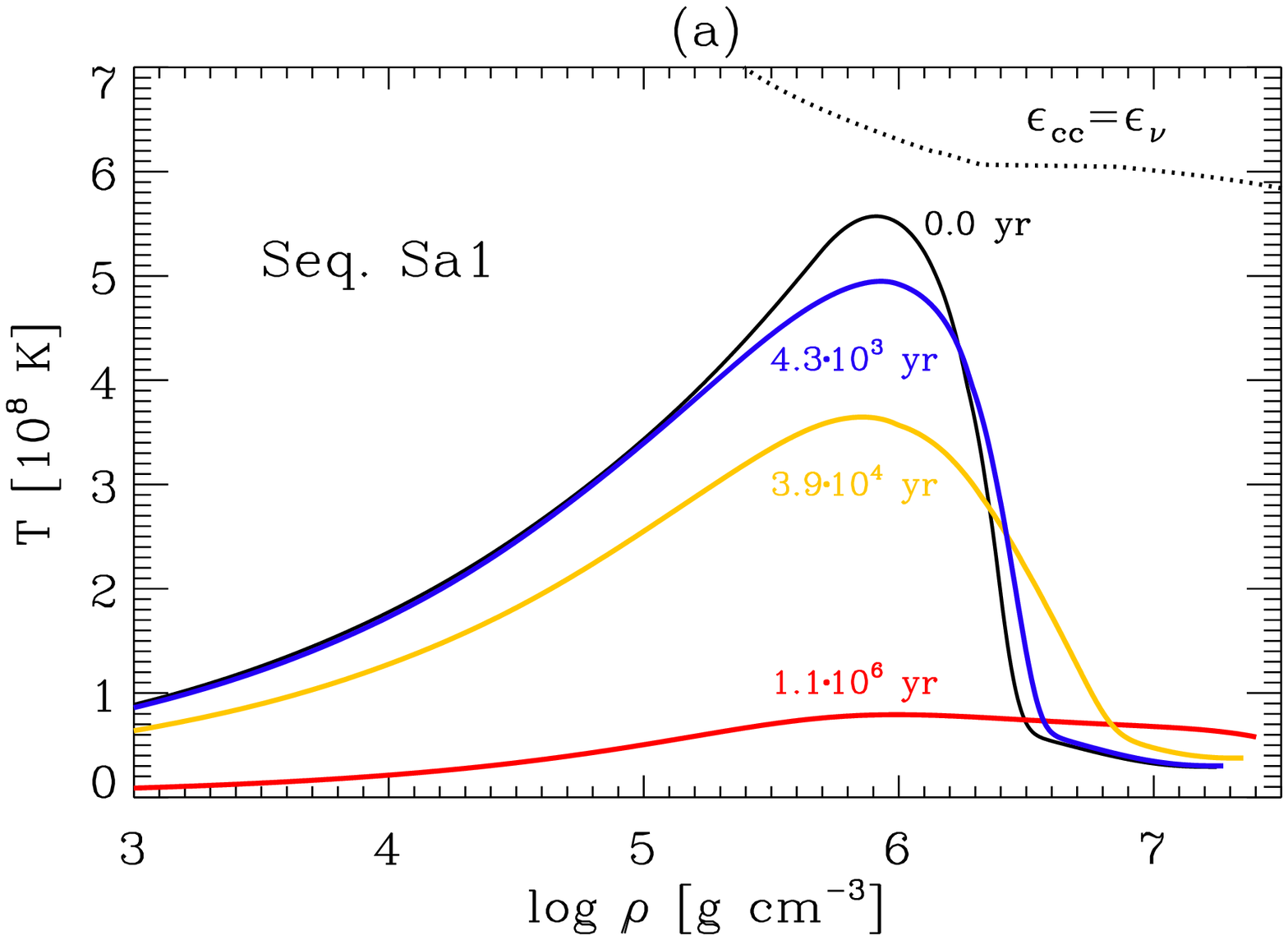}} 
\resizebox{0.4\hsize}{!}{\includegraphics{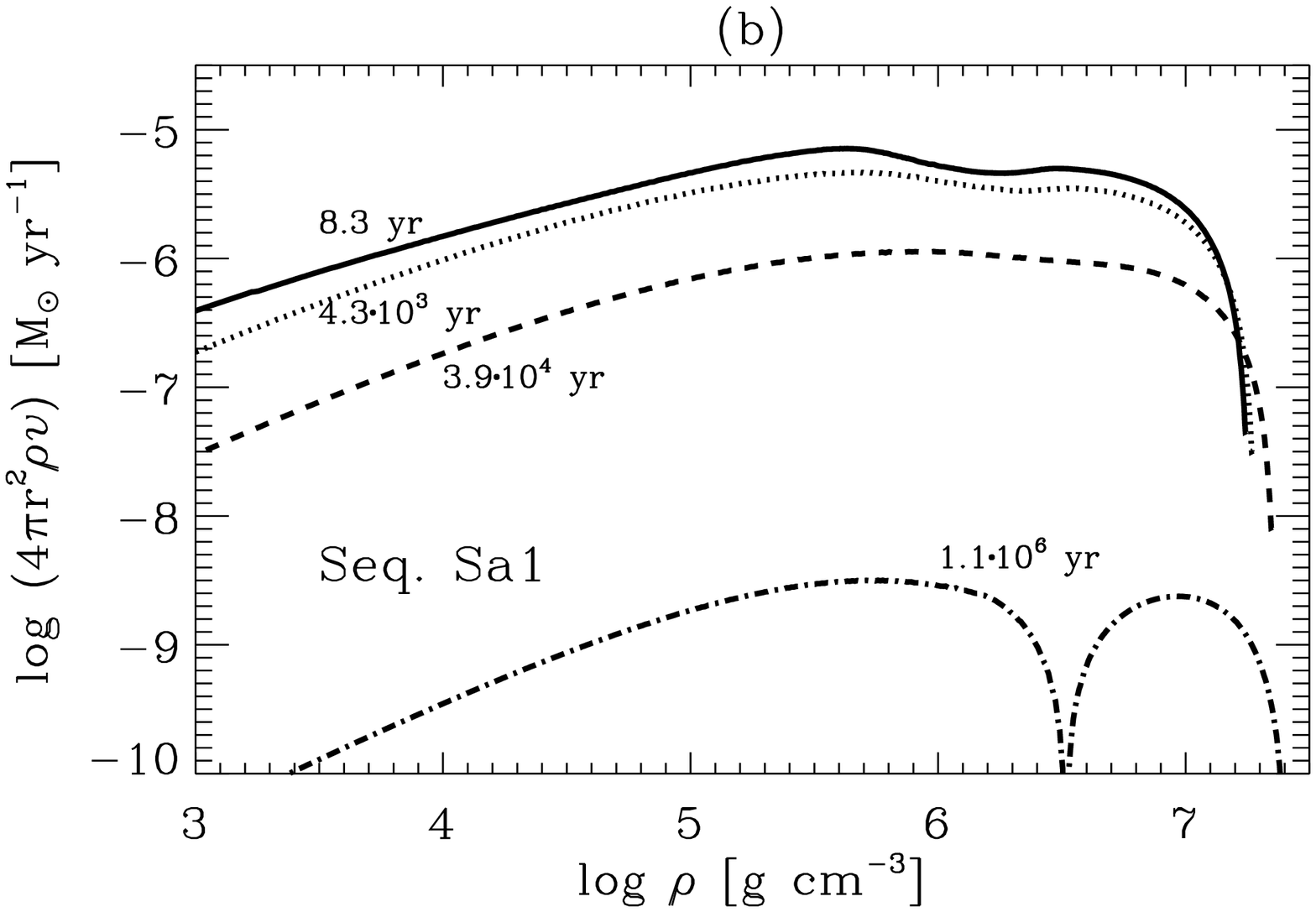}} 
\resizebox{0.4\hsize}{!}{\includegraphics{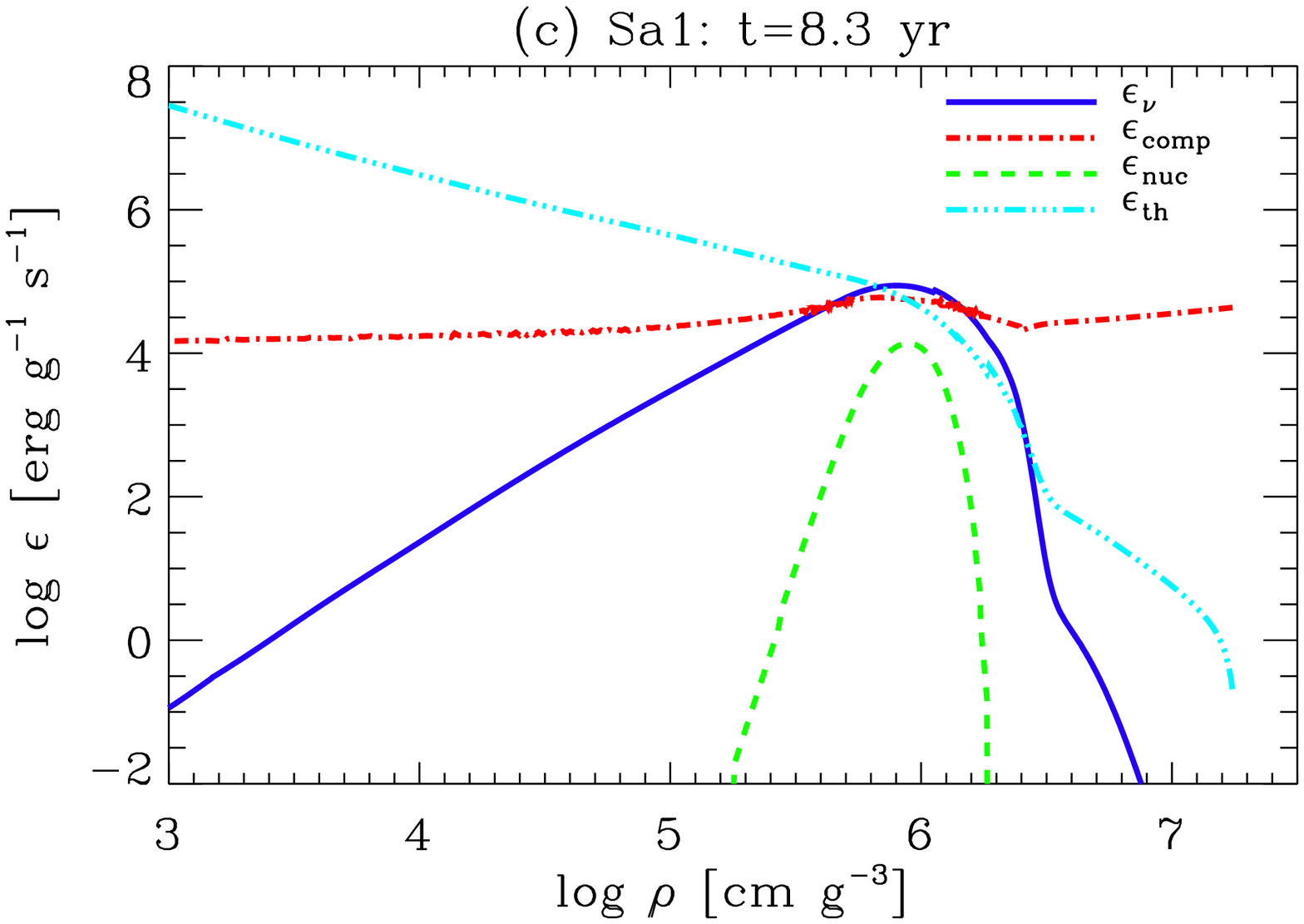}} 
\resizebox{0.4\hsize}{!}{\includegraphics{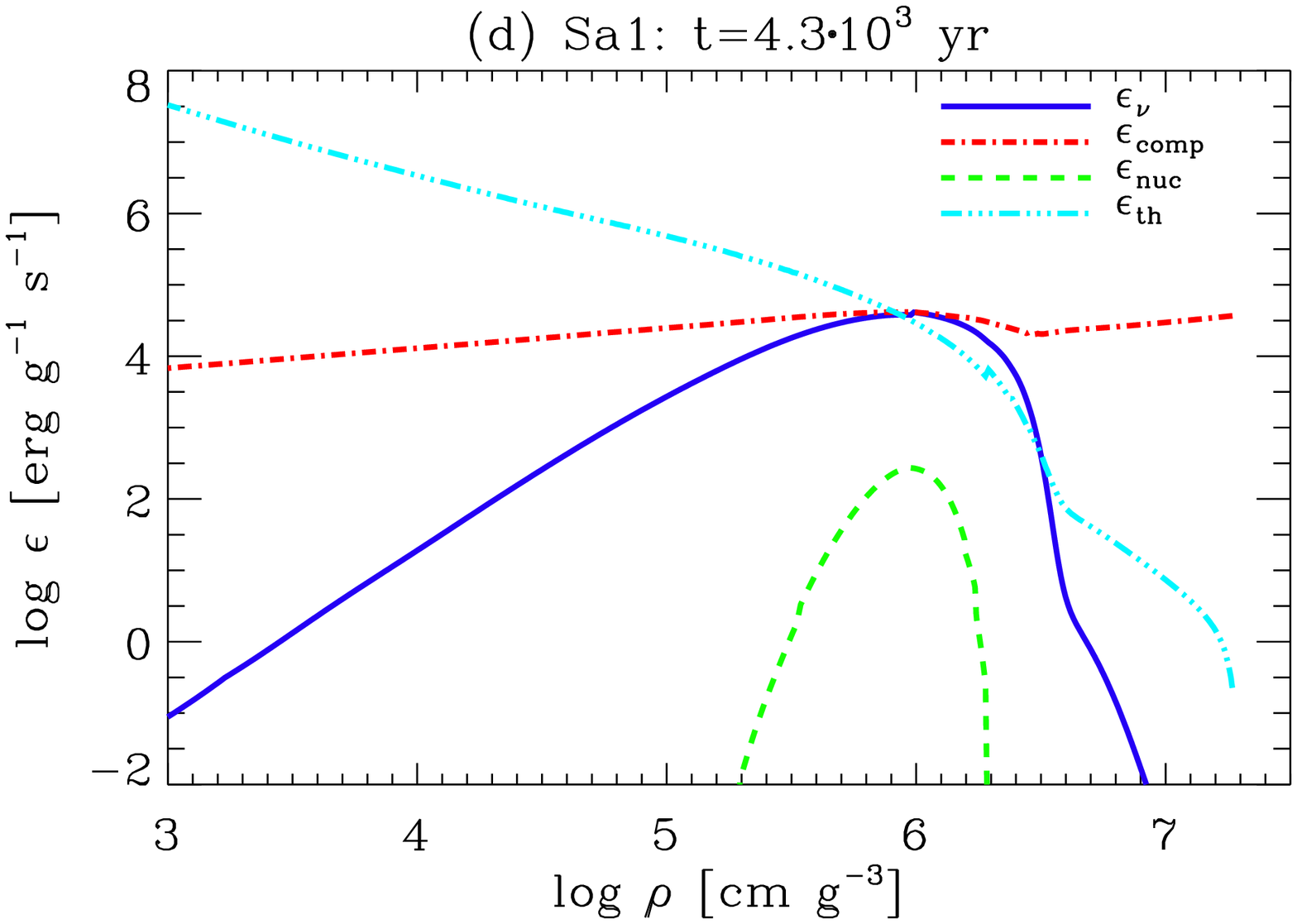}} 
\resizebox{0.4\hsize}{!}{\includegraphics{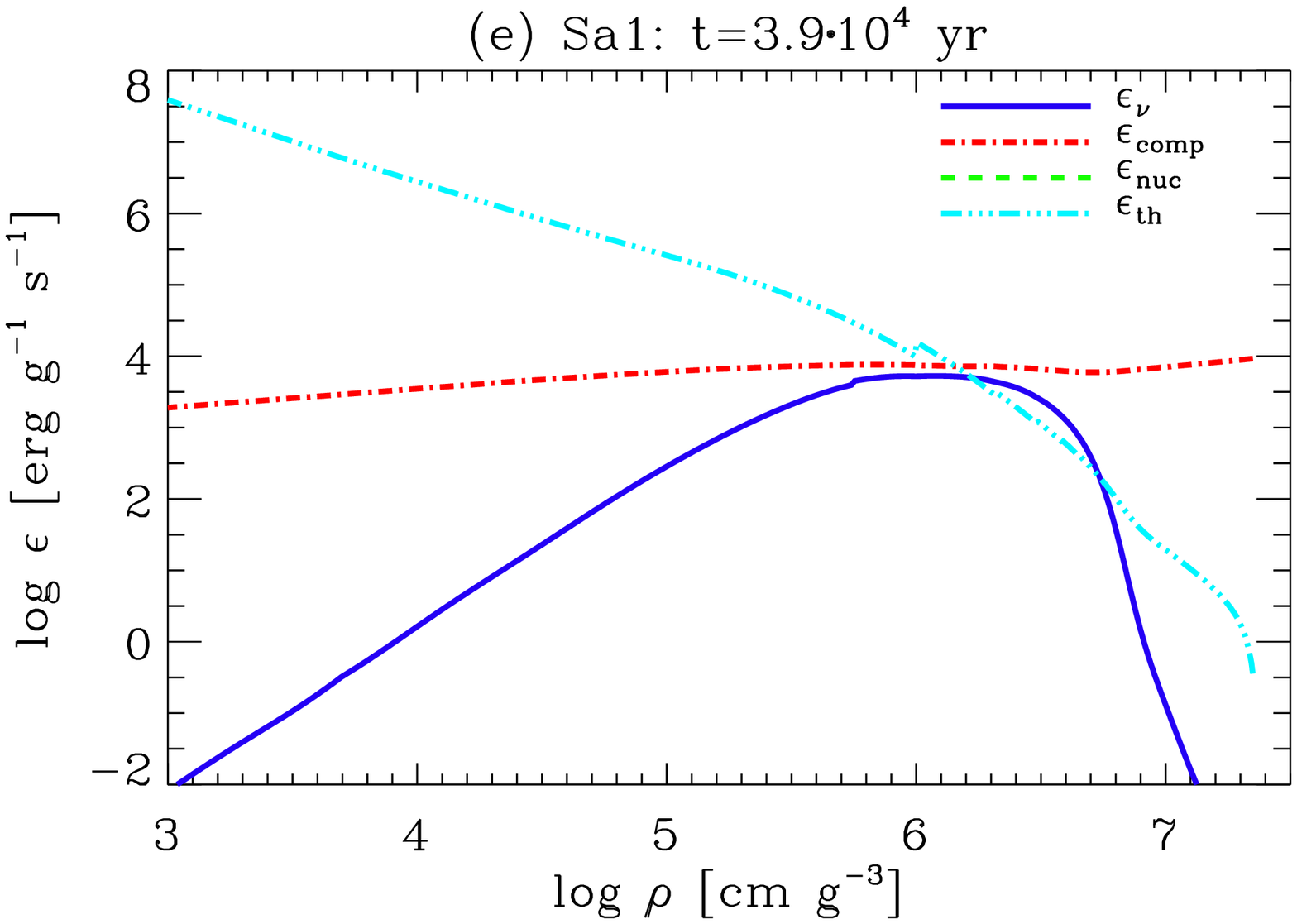}} 
\resizebox{0.4\hsize}{!}{\includegraphics{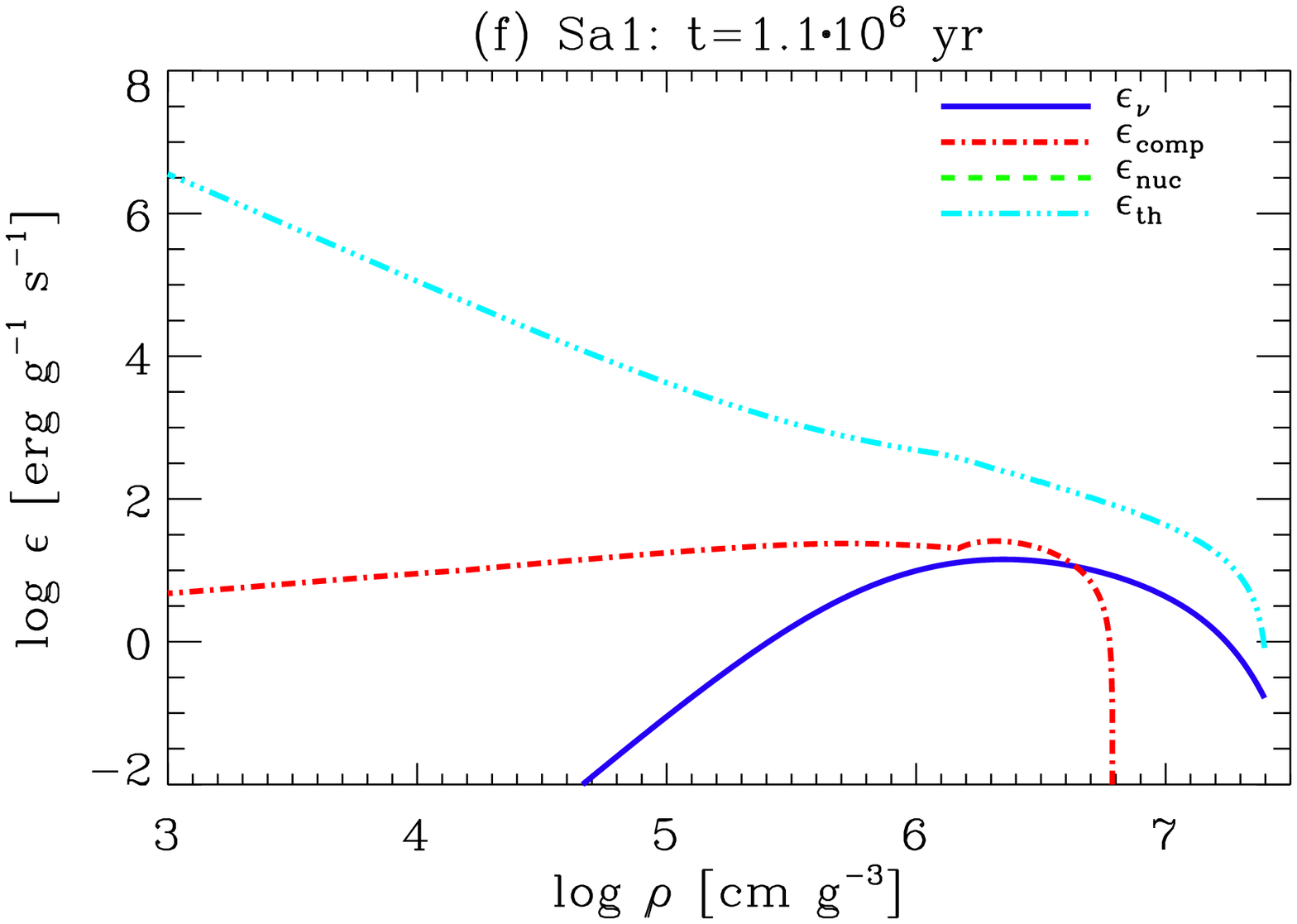}} 
\caption{(a) Evolution of the central remnant in Seq. Sa1 in the $\log 
\rho - T$ plane.  The dotted curve gives the critical temperature 
where the nuclear energy generation rate due to carbon burning equals 
the energy loss rate due to neutrino cooling.  (b) The local effective 
accretion rate ($\dot{M}_\mathrm{eff, r} \equiv 4\pi r^2\rho 
v_\mathrm{r}$) as a function of density in the merger remnant model of 
Seq. Sa1, at different evolutionary epochs as indicated by the labels. 
(c) -- (f) The rates of energy loss/gain due to neutrino 
($\epsilon_{\nu}$) cooling, compressional heating 
($\epsilon_\mathrm{comp}$), nuclear energy generation 
($\epsilon_\mathrm{comp}$) and thermal diffusion 
($\epsilon_\mathrm{th}$) as a function of density in the central 
remnant models of Seq. Sa1 at different evolutionary epochs. Note that 
here $\epsilon_{\nu}$, $\epsilon_\mathrm{comp}$ and 
$\epsilon_\mathrm{nuc}$ represent the values which are used in the 
evolutionary calculations, while $\epsilon_\mathrm{th}$ is an 
order-of-magnitude estimate according to Eq.~(7).  }\label{figSa1} 
\end{center} 
\end{figure*}

As Table~\ref{tabacc} shows, and consistent with the findings of 
\citet{Nomoto85}, such off-center carbon flashes occur regardless of 
the initial mass of the white dwarf, if $\dot{M}_\mathrm{acc} \approx 
10^{-5}~\mathrm{M_\odot~yr^{-1}}$. The results with models including 
rotation show that carbon ignition may be delayed if the effect of 
rotation is included (Table~\ref{tabacc}; see also 
\citealt{Piersanti03a} and \citealt{Saio04}).  The reason is that the 
local effective mass accretion rate ($\dot{M}_\mathrm{eff, r} \equiv 
4\pi r^2\rho v$) inside the white dwarf at a given mass is lower 
because of the centrifugal force.  For instance, in Seq. N0.9, we have 
$\dot{M}_\mathrm{eff, r} \approx 10^{-5}~\mathrm{M_\odot~yr^{-1}}$ at 
around $\rho = 5\times10^5~\mathrm{g~cm^{-3}}$ when 
$t\simeq10^4~\mathrm{yr}$ (Fig.~\ref{figN0.9}b), but 
$\dot{M}_\mathrm{eff, r}$ is lowered by a factor of two in the 
corresponding rotating model at a similar epoch (i.e., 
$\dot{M}_\mathrm{eff, r}\approx 
5\times10^{-6}~\mathrm{M_\odot~yr^{-1}}$), as revealed in 
Fig.~\ref{figmdotR0.9}.  However, carbon ignition occurs well before 
the white dwarf reaches the Chandrasekhar limit, in all model 
sequences considered. Thus, rotation by itself cannot change the 
conclusion of the previous work that the coalescence of double CO 
white dwarfs should lead to accretion-induced collapse rather than a 
thermonuclear explosion, unless the accretion rate is significantly 
lowered, as was also shown by \citet{Piersanti03a} and \citet{Saio04}.

\subsubsection[]{Sequences without angular-momentum loss and mass accretion} 
 
Having understood the physics of the thermal evolution of CO white 
dwarfs which accrete cold matter with a rate close to the Eddington 
limit, we now investigate the evolution of the central remnant model 
consisting of a cold core and a hot envelope as described in 
Sect.~\ref{sectmethod}.  First, we examine the results of the model 
sequences where both angular-momentum loss and mass accretion from the 
Keplerian disc are neglected (i.e., $\tau_\mathrm{J} = \infty$ and 
$\dot{M} = 0$; Seqs~Sa1, Aa1, Ab1, Ac1, Ad1, Ae1, Ba1, \& Ta1). 
 
Fig.~\ref{figSa1}a illustrates the evolution of the central remnant 
for $M_\mathrm{CR} = 1.10~\mathrm{M_\odot}$ in Seq.~Sa1 in the density 
-- temperature plane.  Note that the local peak of temperature at $t = 
0.0$ ($T_\mathrm{p} = 5.6\times10^8~\mathrm{K}$) is significantly 
below the critical temperature for carbon ignition ($T_\mathrm{C-ig}$; 
dotted curve in Fig~\ref{figSa1}a).  It is shown in Fig.~\ref{figSa1}b 
that the local effective accretion rate ($\dot{M}_\mathrm{eff,r}$) 
remains relatively high ($5\times 10^{-6} - 
10^{-5}~\mathrm{M_\odot~yr^{-1}}$) around $\rho = 
10^6~\mathrm{g~cm^{-3}}$, where the local peak of temperature is 
located, for about 5000 yrs. Despite such high effective accretion 
rates, the temperature peak continuously decreases, although the inner 
core becomes somewhat hotter due to compression, and the central 
remnant finally becomes a cold white dwarf.  A few remarkable 
differences compared to the standard accreting white dwarf models are 
found in this regard.  Firstly, since the envelope is very hot, 
neutrino cooling -- in particular by photoneutrinos -- is significant 
from the beginning, and even dominant over the thermal diffusion at 
the interface between the core and the envelope as shown in 
Fig.~\ref{figSa1}c.  In cold-matter accreting white dwarfs, neutrino 
cooling becomes important only after a significant amount of mass has 
been accreted (Fig.~\ref{figN0.9}).  Secondly, the compressional 
heating rate is slightly lower than the neutrino cooling rate around 
the local peak of temperature.  As the contraction of the central 
remnant is mainly determined by the thermal evolution of the envelope, 
the local accretion rate is in fact controlled by the cooling process. 
This explains why we have $\epsilon_\mathrm{comp} \approx 
\epsilon_\nu$ around the local peak of temperature for the initial 
$\sim 10^4$ yrs, and why the local peak of temperature continuously 
decreases despite the relatively high effective accretion rate.  This 
conclusion is the same for all other sequences with a $T_\mathrm{P}$ 
that is significantly lower than $T_\mathrm{C-ig}$ (Seq.~Aa1, Ab1, \& 
Ba1) including the non-rotating case (Seq.~Ta1; Table~\ref{tabmerger}).

\begin{figure} 
\begin{center} 
\resizebox{0.9\hsize}{!}{\includegraphics{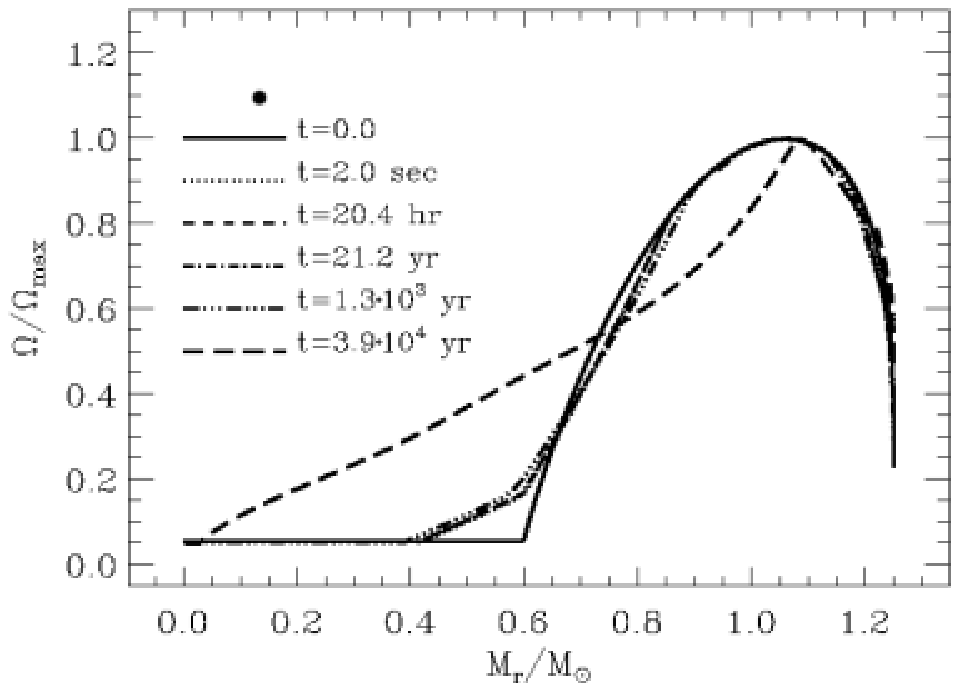}} 
\resizebox{0.9\hsize}{!}{\includegraphics{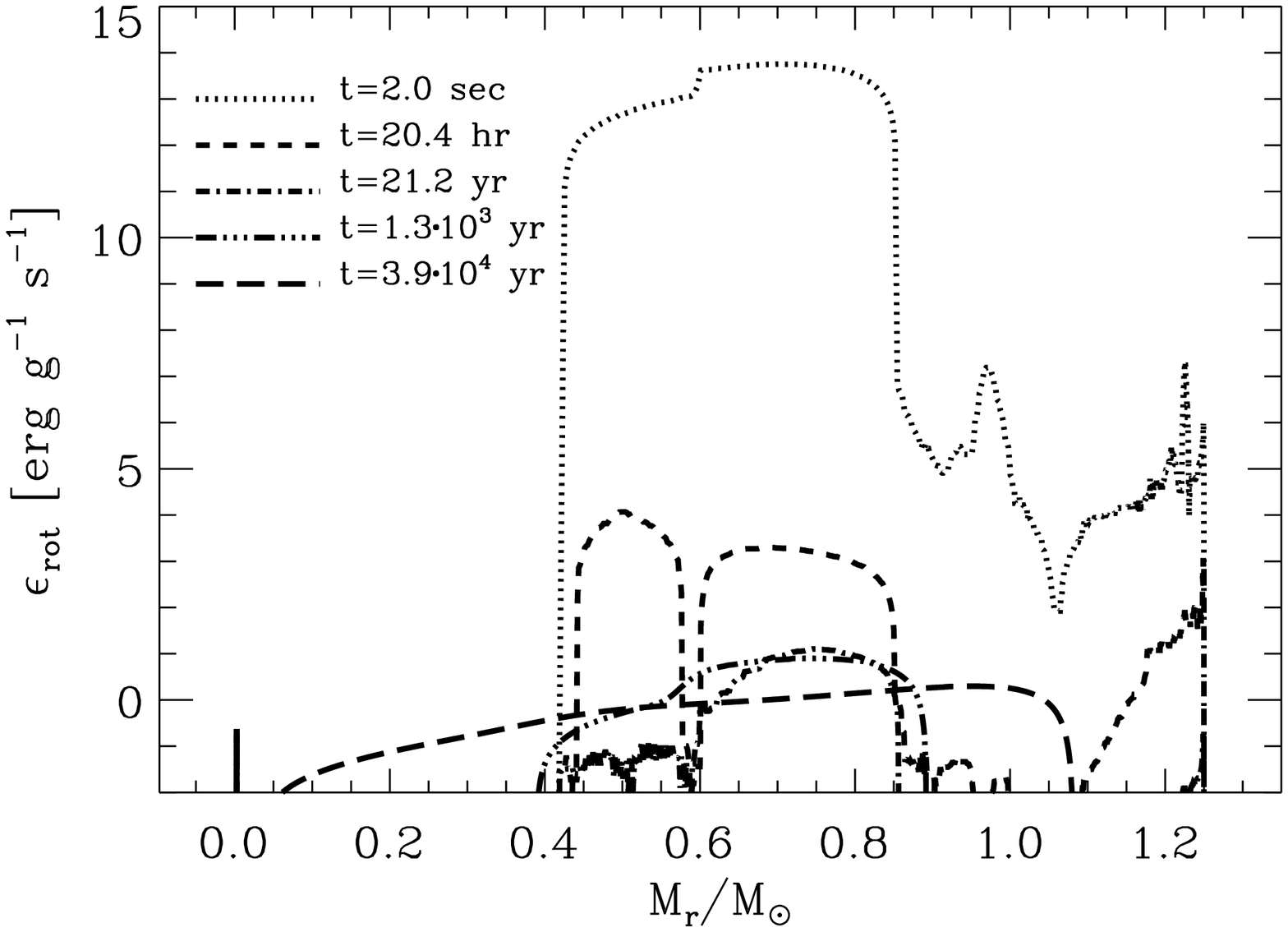}} 
\caption{\emph{Upper panel}: The angular velocity relative to the local 
maximum as a function of the mass coordinate in the central remnant 
models of Seq. Aa1 at different evolutionary epochs. \emph{Lower 
panel}: the rate of rotational energy dissipation 
($\epsilon_\mathrm{rot}$; see YL04) as a function of the 
mass coordinate in the corresponding models shown in the upper panel. 
}\label{figomegAa} 
\end{center} 
\end{figure}

We find that, in Seq.~Sa1, the differentially rotating layers at the 
interface between the core and the envelope in the initial central 
remnant model are stable against the dynamical shear instability 
(DSI).  They are, however, unstable to the DSI in other sequences, 
where the interface is more degenerate (see YL04 for discussions on 
the DSI).  Consequently, in Seq.~Aa1 for example, the rate of 
rotational energy dissipation ($\epsilon_\mathrm{rot}$) appears to be 
very high initially (Fig.~\ref{figomegAa}).  The differentially 
rotating layers are rapidly smeared out by the dynamical shear 
instability (see the discussion in Sect.~2 in YL04), and 
$\epsilon_\mathrm{rot}$ falls below the thermal diffusion and/or 
neutrino cooling rate only within 20 yrs.  Hence we conclude that the 
rotational energy dissipation does not play an important role for the 
long-term evolution of the central remnant.

\begin{figure} 
\begin{center} 
\resizebox{0.9\hsize}{!}{\includegraphics{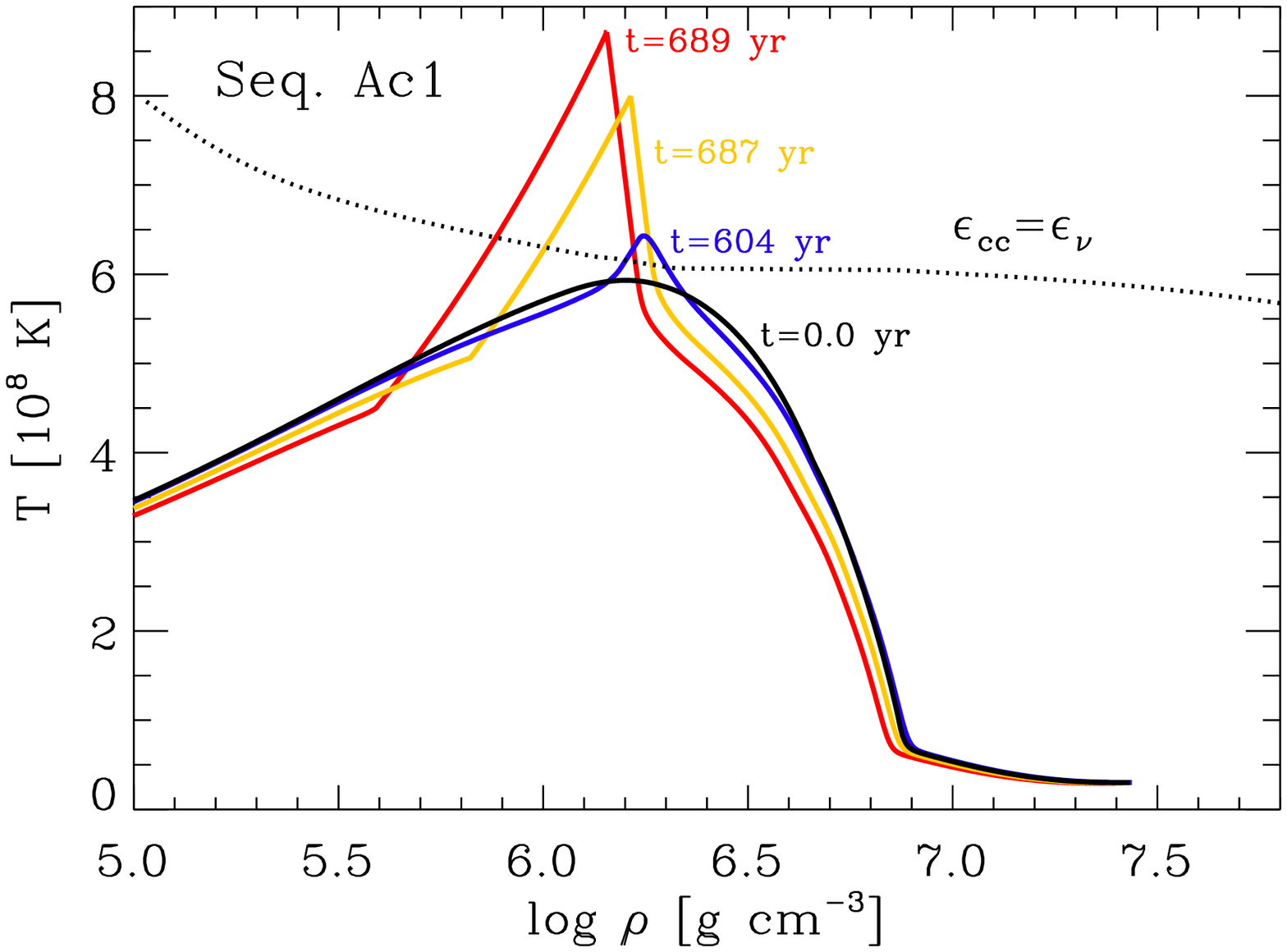}} 
\resizebox{0.9\hsize}{!}{\includegraphics{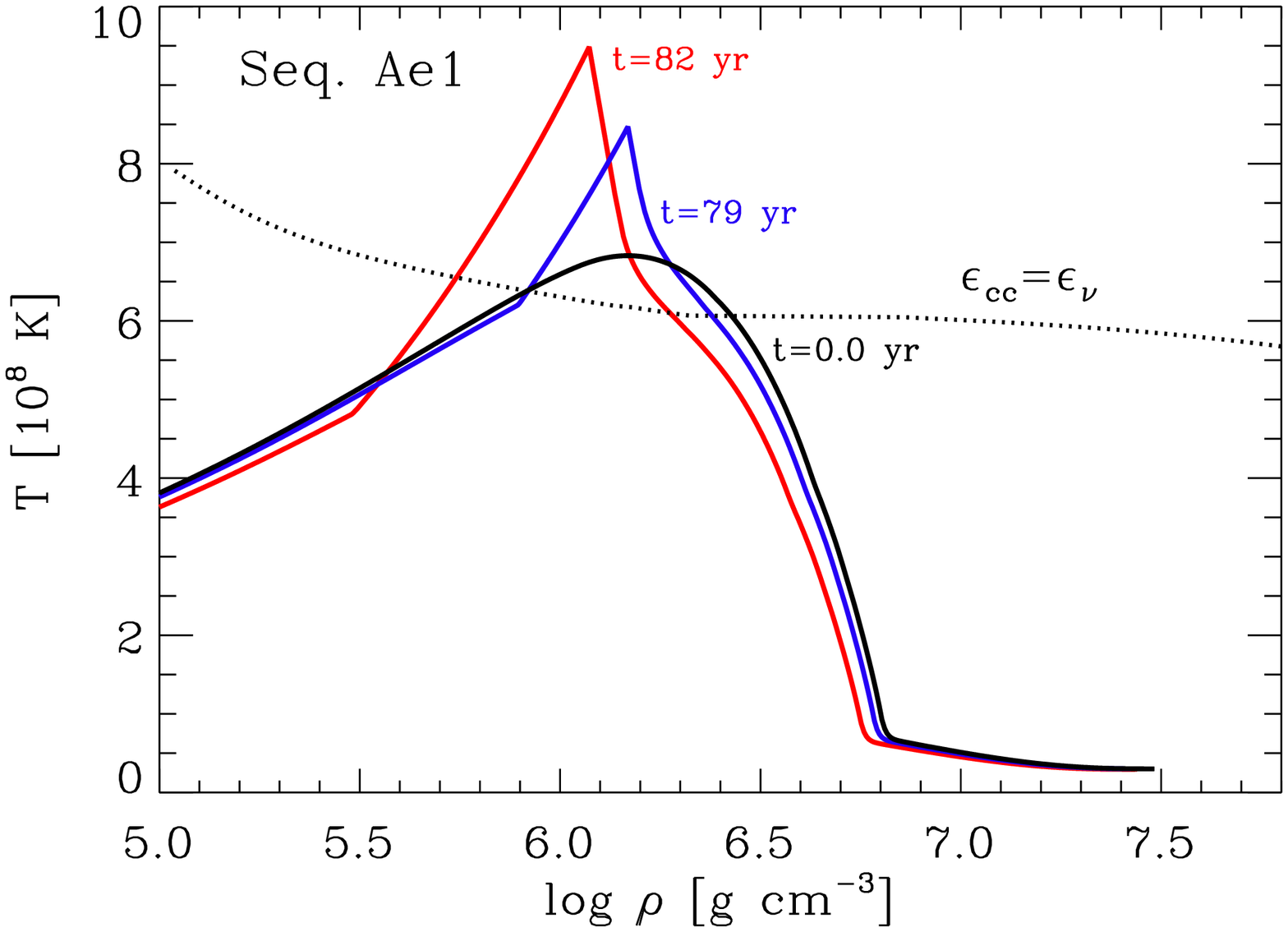}} 
\caption{Evolution of the central remnant in Seqs~Ac1 (upper panel) 
and Ae1 (lower panel), in the $\log \rho - T$ plane.  The dotted curve 
gives the critical temperature where nuclear energy generation rate 
due to carbon burning equals to energy loss rate due to neutrino 
cooling.  }\label{figrhotAc1} 
\end{center} 
\end{figure}

Fig.~\ref{figrhotAc1} shows how the evolution of the central remnant 
changes if the local peak of temperature in the initial model 
($T_\mathrm{p}$) is close to or above the critical limit for carbon 
burning ($T_\mathrm{C-ig}$), with Seqs~Ac1 and Ae1 as examples.  In 
contrast to Seq.\ Sa1 or Aa1, carbon burning dominates the evolution 
very soon in both sequences, and the temperature increases rapidly. 
Although the further evolution has not been followed in the present 
study, it is most likely that the carbon-burning flame propagates 
inward such that the central remnant is converted into an 
ONeMg white dwarf within several thousand years as 
shown by \citet{Saio98}. 
 
As summarized in Table~\ref{tabmerger}, all other sequences follow the 
same evolutionary pattern: off-center carbon ignition is avoided in 
Seqs Ab1, Ba1, and Ta1 where $T_\mathrm{p}$ is significantly below 
$T_\mathrm{C-ig}$, while carbon ignites off-center in the other 
sequences where $T_\mathrm{p} \ga T_\mathrm{C-ig}$.  It is thus 
remarkable that the thermal evolution of the central remnant is 
sensitively determined by the local peak of temperature in the 
quasi-static equilibrium state. 
 
In conclusion, in the absence of angular-momentum loss and mass 
accretion from the Keplerian disc, the thermal evolution of the 
central remnant is roughly controlled by neutrino cooling at the 
interface between the core and the envelope, and off-center carbon 
burning may be avoided as long as $T_\mathrm{p} < T_\mathrm{C-ig}$, 
while it seems inevitable if $T_\mathrm{p} \ga T_\mathrm{C-ig}$.

\subsection[]{Effect of angular momentum loss}

\begin{figure*} 
\begin{center} 
\resizebox{0.4\hsize}{!}{\includegraphics{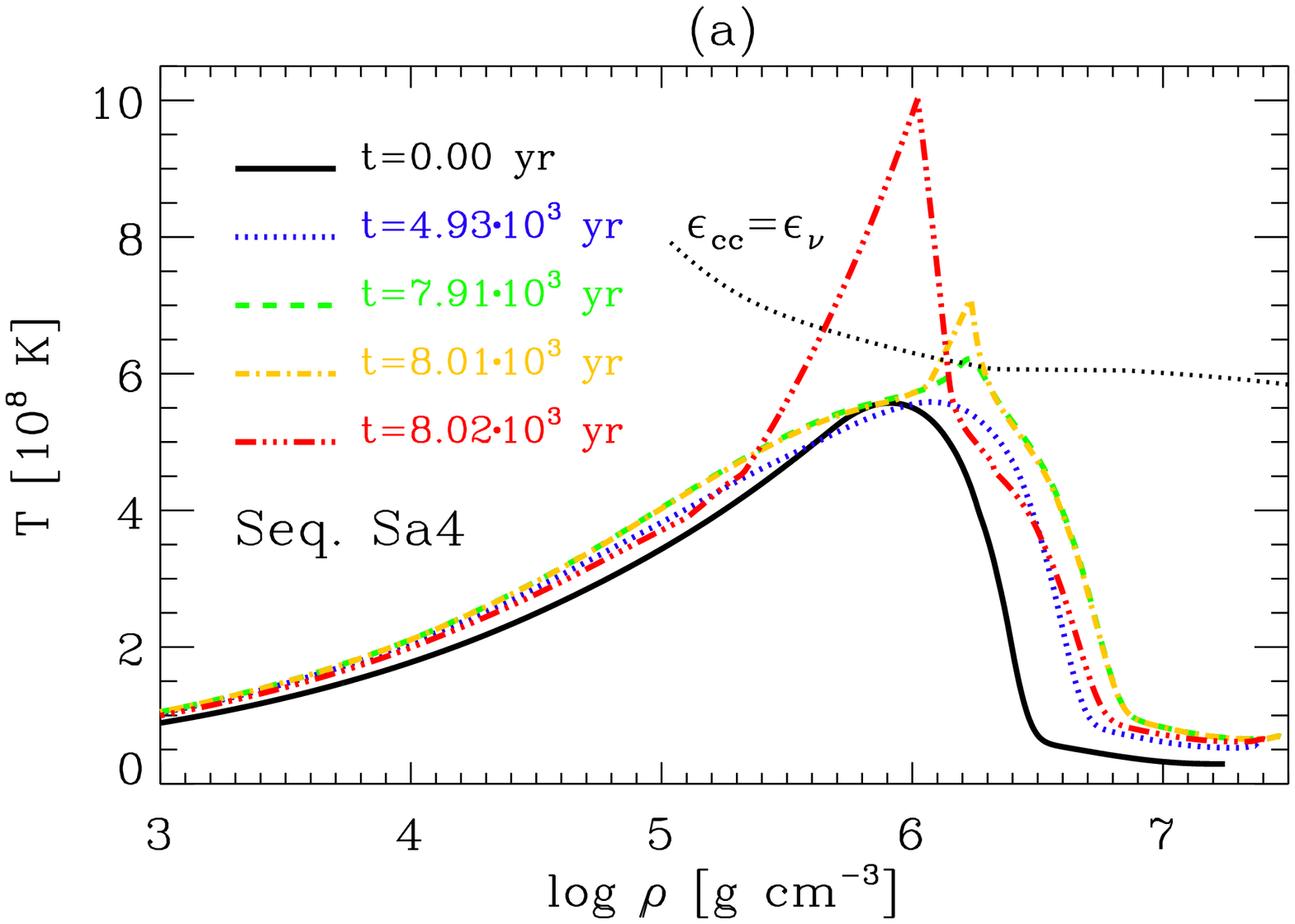}} 
\resizebox{0.4\hsize}{!}{\includegraphics{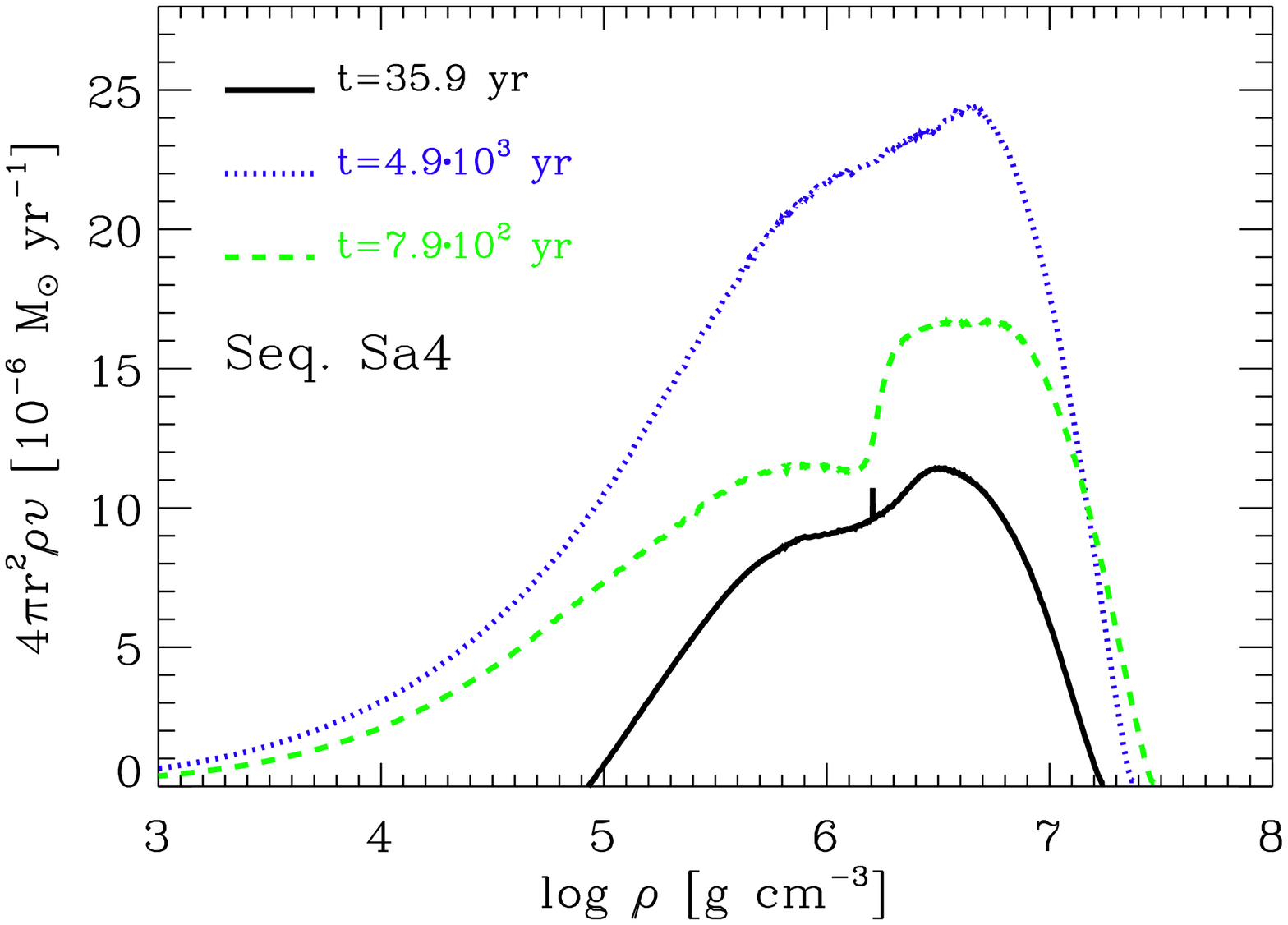}} 
\resizebox{0.4\hsize}{!}{\includegraphics{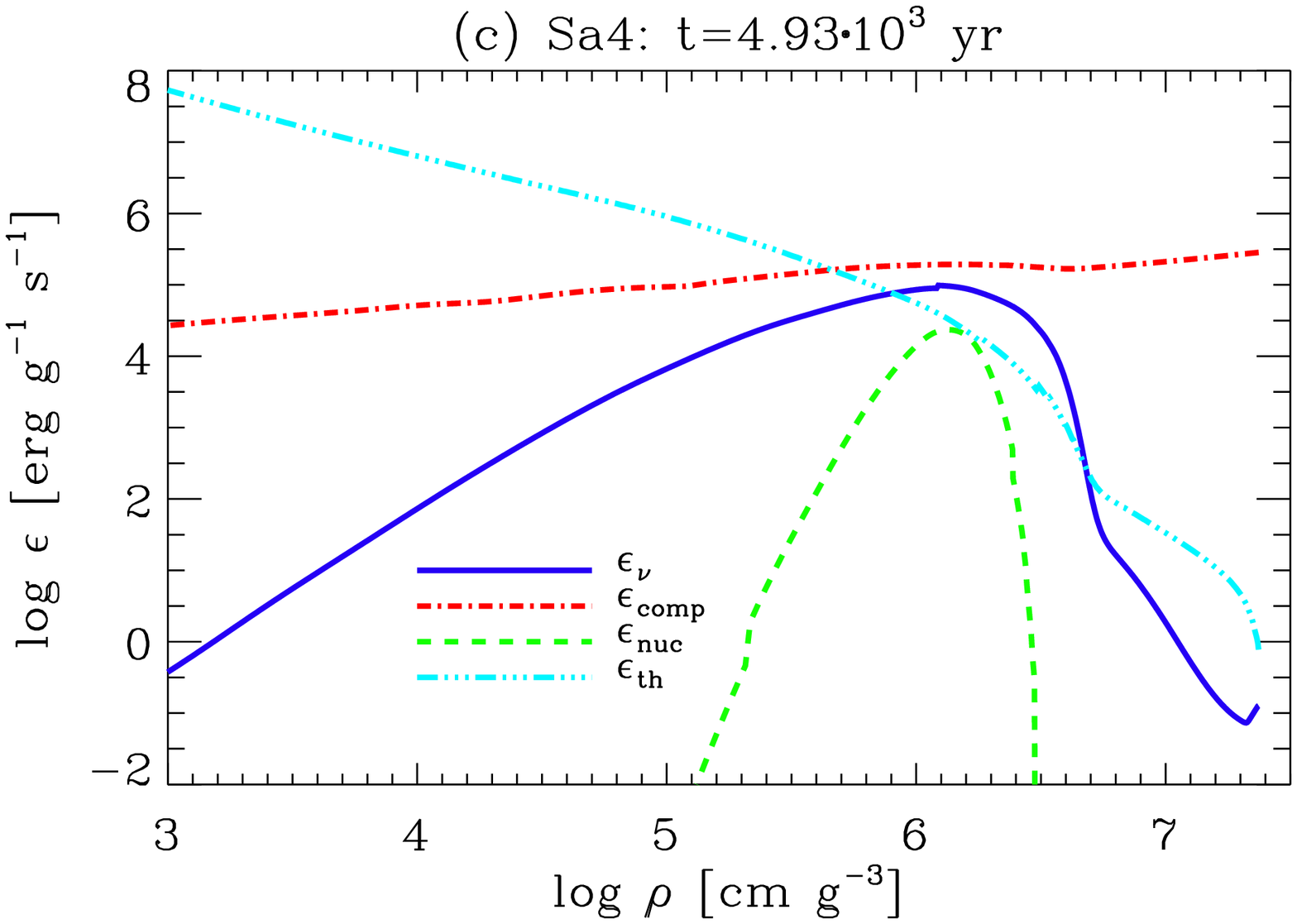}} 
\resizebox{0.4\hsize}{!}{\includegraphics{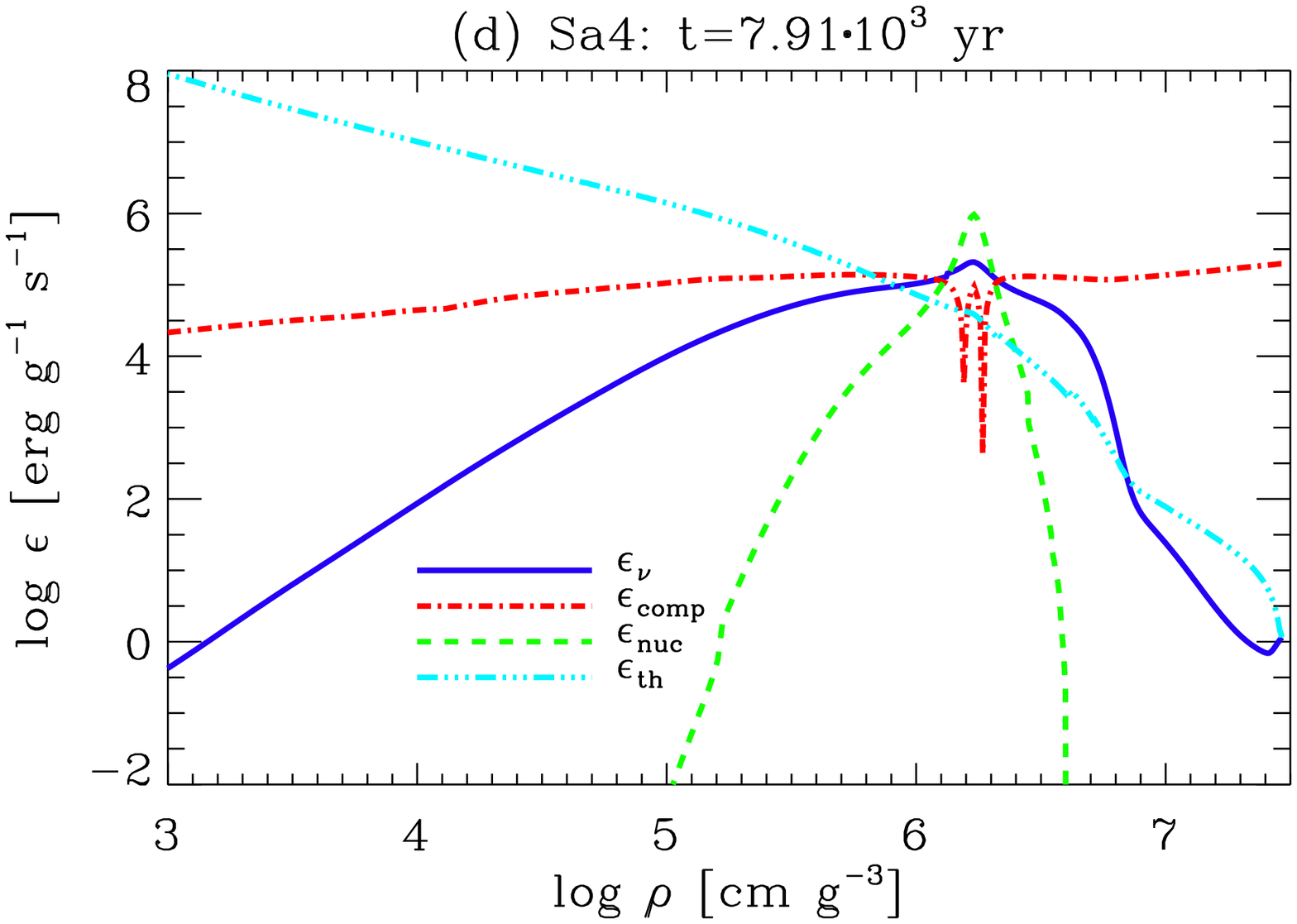}} 
\caption{Same as in Fig.~\ref{figSa1}, but for Seq.~Sa4. 
 }\label{figSa4} 
\end{center} 
\end{figure*}

In Seqs~Sa2 -- Sa5, the central remnant has the same initial 
conditions as in Seq.~Sa1, angular momentum loss from the white dwarf 
with different time scales $\tau_\mathrm{J}$ is considered according to Eq.~(4). 
Note that off-center carbon ignition occurs in Seqs Sa2, Sa3 \& Sa4, 
where $\tau_\mathrm{J} \la 10^4~\mathrm{yr}$, while it is avoided in 
Seq. Sa5 where $\tau_\mathrm{J} = 10^5~\mathrm{yr}$.  These results 
indicate that off-center carbon ignition should be induced if the 
angular-momentum loss occurs too rapidly for neutrino cooling or 
thermal diffusion to control the effective mass accretion.  For 
instance, Fig.~\ref{figSa4} shows that in Seq.~Sa4, where 
$\tau_\mathrm{J}=10^4~\mathrm{yr}$, the effective mass accretion rate 
reaches a few $10^{-5}~\mathrm{M_\odot~yr^{-1}}$ at the interface 
between the core and the envelope ($\rho \approx 
10^6~\mathrm{g~cm^{-3}}$), and the compressional heating rate exceeds 
the neutrino cooling rate.

\begin{figure} 
\begin{center} 
\resizebox{\hsize}{!}{\includegraphics{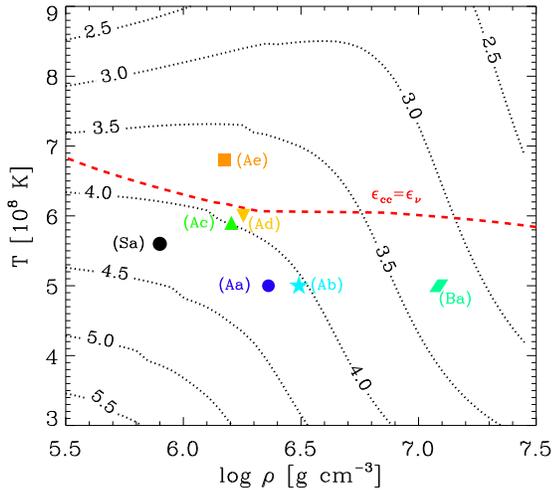}} 
\caption{Contour lines of the neutrino cooling time scale ($\tau_\nu\equiv 
TC_\mathrm{p}/\epsilon_\nu$) in the $\log \rho - T$ plane. The level 
at each line gives $\log \tau_\nu$ in units of years.  The dashed 
line denotes the critical temperature for carbon ignition.  The local 
peak of temperature and the corresponding density in the initial model 
of Seqs  Sa, Aa, Ab, Ac, Ae, and Ba are marked by the filled symbols 
as indicated by the labels.  }\label{conttaunu} 
\end{center} 
\end{figure}

It is shown that the critical angular-momentum-loss time scale, 
$\tau_\mathrm{J}$, for off-center carbon ignition ($\tau_\mathrm{J, 
crit}$) is smaller for Seqs Ab and Ba than for Seqs Sa and Aa: 
$\tau_\mathrm{J, crit} \approx 10^3$ for Seqs Ab and Ba, and 
$\tau_\mathrm{J, crit} \approx 10^4$ for Seqs Sa and Aa.  This is due 
to the different local thermodynamic properties at the interface 
between the core and the envelope in different central remnant models. 
As shown in Fig.~\ref{conttaunu}, higher density and/or temperature at 
the interface result in a shorter neutrino cooling time, making it 
possible to avoid local heating for a smaller $\tau_\mathrm{J}$.  In 
other words, $\tau_\mathrm{J, crit}$ roughly corresponds to the time 
scale for neutrino cooling at the local peak of temperature 
($\tau_\mathrm{\nu,p}$). 
 
From this experiment, we conclude that, in the absence of mass accretion from the Keplerian disc,  
carbon ignition  may be avoided in the central remnant,  
if $T_\mathrm{max,init} < T_\mathrm{C-ig}$, and if $\tau_\mathrm{J} > \tau_\mathrm{\nu,p}$.

\subsection[]{Mass accretion from the Keplerian disc}\label{acc} 
 
In reality, mass accretion from the Keplerian disc onto the central 
remnant is expected.  The accretion rate is determined by the 
viscosity of the disc, which is not well known.  However, we expect 
the accretion rate from a Keplerian disc may be significantly lower 
than from a pressure-supported thick disc that was assumed in previous 
studies.  Our results, as summarized in Table~\ref{tabmerger}, indicate 
that even with mass accretion, the central remnant with $T_\mathrm{p} 
< T_\mathrm{C-ig}$ can avoid off-center carbon ignition if the 
accretion rate is sufficiently low (i.e., $\dot{M} < 
5\times10^{-6}...10^{-5}~\mathrm{M_\odot~yr^{-1}}$), and if 
$\tau_\mathrm{J} > \tau_\mathrm{\nu,p}$ (see Table~\ref{tabmerger}). 
 
The thermal history of the central remnant in those sequences where 
carbon ignites off-center is similar to that of the white dwarf in 
classical accretion model sequences.  However, as the central remnant 
has a rapidly rotating hot envelope, carbon ignition is significantly 
delayed compared to the case of classical accretion.  In Seq.~N1.2, 
where $M_\mathrm{init} = 1.2~\mathrm{M_\odot}$ and $\dot{M}_\mathrm{acc} = 
10^{-5}~\mathrm{M_\odot~yr^{-1}}$, carbon ignites only when about 
$0.025~\mathrm{M_\odot}$ is accreted, while in Seq.~Aa7 more than 
$0.15~\mathrm{M_\odot}$ have to be accreted to induce carbon ignition 
at the same accretion rate, despite its higher initial mass.  On the 
other hand, the comparison of Seq. Aa7 with Seq. Aa10 indicates that 
off-center carbon ignition is delayed if the central remnant keeps 
more angular momentum.  The critical accretion rate for inducing 
off-center carbon ignition is thus difficult to precisely determine, 
as our 1-D models significantly underestimate the effect of the 
centrifugal force, especially in the envelope where carbon ignites. 
In addition, the physics of angular momentum loss/gain is not well 
understood yet, as discussed in \citet{Yoon05}. 
 
Note that $M_\mathrm{WD,ig}$ in Seqs  Aa10, Ba6 and Ba7 is already 
very close, or even above the Chandrasekhar limit. However, the 
central density in those models is still smaller by an order of 
magnitude than the critical limit for carbon ignition due to the 
effect of rotation. As the carbon-burning flame will propagate inwards 
within several thousand years \citep{Saio98}, only about $\sim 
0.05~\mathrm{M_\odot}$ may be further accreted by the time the burning 
flame reaches the center, and the central density may not become high 
enough to induce a thermonuclear explosion before the whole central 
remnant is converted into an ONeMg white dwarf.   
(Super-) Chandrasekhar mass ONeMg white dwarfs produced in this way 
will eventually collapse to a neutron star (see \citealt{Yoon05}; 
\citealt{Dessart06}).

On the other hand, the white dwarf continuously grows to/above the 
Chandrasekhar limit ($\approx 1.4~\mathrm{M_\odot}$) without suffering 
carbon ignition (neither at the center nor off-center) in Seqs Sa8, 
Sa9, Sa10, Sa11, Aa8, Aa9, Ab6, and Ba8.  The outcome in these cases 
is thus the formation of a (super-) Chandrasekhar mass CO white dwarf, 
which will eventually explode as a Type Ia supernova.
The mass of the exploding white dwarf should 
depend on the amount of angular momentum \citep{Yoon05} and cannot 
exceed the mass budget of merging white dwarfs.  Fig.~\ref{figmj} 
shows the evolutionary paths of the central remnant for Seqs Sa8, Sa9 
\& Sa11 as examples in the mass -- angular momentum plane. Note that 
the central remnant initially has a large amount of angular momentum 
($J = 1.11\times10^{50}~\mathrm{erg~s}$), such that without loss/gain 
of angular momentum, it should accrete matter until it reaches 
$M\simeq1.68~\mathrm{M_\odot}$ where it explodes in a SN Ia explosion. 
In Seqs~Sa8 and Sa9, the accretion time scale ($\tau_\mathrm{acc}$) is 
longer than the angular momentum loss time scale, and the total 
angular momentum of the white dwarf continuously decreases while the 
total mass increases. Consequently, carbon ignites at the center when 
the white dwarf grows to $1.50~\mathrm{M_\odot}$ and 
$1.42~\mathrm{M_\odot}$ for Seqs~Sa8 and Sa9, respectively. In 
Seq.~Sa11, on the other hand, both mass and angular momentum of the 
central remnant continuously increase, given that $\tau_\mathrm{acc} 
\la \tau_\mathrm{J}$, and a SN Ia explosion is expected only when 
$M\simeq1.70~\mathrm{M_\odot}$.  Note that this is even larger than 
the mass budget of the binary system considered for this sequence 
(i.e, $0.9~\mathrm{M_\odot}+0.6~\mathrm{M_\odot}$). In nature, the 
white dwarf must stop growing in mass when $M = 1.5~\mathrm{M_\odot}$, 
and a SN Ia explosion will be induced only when a sufficient amount of 
angular momentum has been removed, e.g. via gravitational wave 
radiation, as illustrated by the path {Sa11-B} in Fig.~\ref{figmj}. 

\section[]{Conclusion and Discussion}\label{sectdiscussion}

We have explored the dynamical and secular evolution of the merger of 
double CO white dwarf binaries whose total mass exceeds the 
Chandrasekhar limit.  Based on our new SPH simulation of the 
coalescence of two CO white dwarfs of $0.9~\mathrm{M_\odot}$ and 
$0.6~\mathrm{M_\odot}$, we suggest that the immediate post-merger 
remnant is best described as a differentially rotating CO star 
consisting of a slowly rotating cold core and a rapidly rotating hot 
envelope that is surrounded by a Keplerian disc rather than as ``cold 
white dwarf + thick disc" system, as in previous investigations.  
The evolution of such a CO star is determined by the thermal evolution 
of the envelope, and the growth of the core is controlled by the cooling due to 
neutrino emission and thermal diffusion, which is fundmentally different 
from the assumption  of ``forced accretion of cold matter". 
 
Our 1-D stellar evolution models of the central remnant, i.e. the cold core and 
the hot envelope, which include the effects of rotation, indicate that 
there are three necessary conditions for the merger remnant to avoid 
off-center carbon ignition such that a SN Ia may be produced:
\begin{enumerate} 
\item The local peak of temperature of the merger remnant at the 
interface between the core and the envelope must be lower than the 
critical temperature for carbon ignition ($T_\mathrm{p} < 
T_\mathrm{C-ig}$). 
\item The time scale for angular-momentum loss from the central remnant by  
 must be larger than the neutrino cooling time scale at the interface 
($\tau_\mathrm{J} > \tau_\mathrm{\nu, P}$).  
\item Mass accretion from the Keplerian disc onto the central remnant 
must be sufficiently slow ($\dot{M}_\mathrm{acc} \la 
5\times10^{-6}...10^{-5}~\mathrm{M_\odot~yr^{-1}}$). 
\end{enumerate} 
Our new SPH simulation confirms that at least the first condition 
($T_\mathrm{p} < T_\mathrm{C-ig}$) should be fulfilled in the 
CO white dwarf binary considered.

As emphasized in Sect.~\ref{sectmethod}, our 1-D models significantly 
underestimate the effect of the centrifugal force on the stellar 
structure in the rapidly rotating outermost layers.  However, since 
thermal diffusion always dominates over both neutrino cooling and 
compressional heating in the outer envelope ($\rho \la 
10^5...10^6~\mathrm{g~cm^{-3}}$) above the interface, as shown in 
Figs.~\ref{figN0.9} and~\ref{figSa4}, the detailed structure of the 
rapidly rotating outermost layers above the interface may not 
significantly affect our results on the thermal evolution of the 
merger remnant, as long as the angular momentum of the envelope is not 
lost faster than the local neutrino cooling time scale at the 
interface.  On the other hand, mass accretion from the Keplerian disc 
should occur preferentially along the equatorial plane of the 
envelope.  As shown in the SPH simulation, the envelope is more 
extended along the equatorial plane, where most angular momentum is 
deposited, than along the polar axis, and the resultant compressional 
heating must be much weakened, compared to the case of our 1-D models. 
The enhanced role of rotation must thus help to increase the critical 
mass accretion rate for inducing off-center carbon ignition, in favor 
of producing a Type Ia supernova. 
 
We have concluded that the loss of angular momentum on a short time 
scale ($\tau_\mathrm{J} \la \tau_\mathrm{\nu, p} \approx 
10^4...10^5~\mathrm{yr}$) may induce off-center carbon ignition even 
when $T_\mathrm{max,init} < T_\mathrm{C-ig}$.  Rapidly rotating 
compact stars may experience loss of angular momentum by gravitational 
wave radiation, due to either the bar-mode instability or the r-mode 
instability.  The onset of the dynamical or secular bar-mode 
instability requires a very high ratio of the rotational energy to the 
gravitational energy: $E_\mathrm{rot}/E_\mathrm{grav} \ga 0.2$ for the 
dynamical bar-node instability, and $E_\mathrm{rot}/E_\mathrm{grav} 
\ga 0.14$ for the secular bar-mode instability 
(e.g. \citealt{Shapiro83}). As both our 1-D models and SPH simulation 
give a value of $E_\mathrm{rot}/E_\mathrm{grav}$ that is much lower 
(about 0.06 -- 0.07) than those critical limits, the bar-mode 
instability may not be relevant.  The r-mode instability may operate, 
in principle, even with such a low $E_\mathrm{rot}/E_\mathrm{grav}$ 
\citep{Andersson98,Friedman98}.  However, we estimate that the growth 
time of the r-mode instability ($\tau_\mathrm{r}$), using our central 
remnant models and following \citet{Lindblom99}, is $\ga 
10^6~\mathrm{yr}$, which is much longer than the local neutrino 
cooling time scale ($\tau_{\nu, \mathrm{p}} \approx 
10^4~\mathrm{yr}$).  Alternatively, angular momentum might be 
transported from the accreting star into the Keplerian disc when the 
accretor reaches critical rotation. Calculations by \citet{Saio04} 
indicate, however, that the decrease of the total angular momentum due 
to such an effect is not significant in accreting white dwarfs.  In 
conclusion, neither gravitational wave radiation nor outward 
angular-momentum transport is likely to lead to a rapid loss of 
angular momentum from the central remnant such that $\tau_\mathrm{J} < 
\tau_\mathrm{\nu, p}$, unless magnetic torques are important. 
 
\begin{figure} 
\begin{center} 
\resizebox{\hsize}{!}{\includegraphics{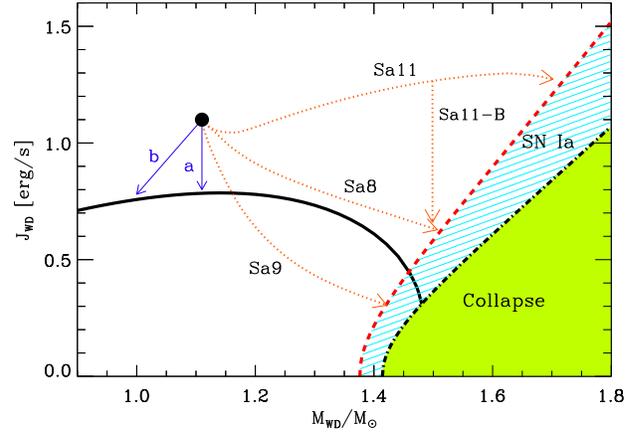}} 
\caption{Evolution of the central remnant in the mass -- angular 
momentum plane. The thick solid curve shows the angular momentum of a 
rigidly rotating white dwarf with critical rotation at the surface as 
a function of the white dwarf mass.  The thick dashed curve and the 
thick dot-dashed curve give the critical angular momentum for a 
differentially rotating CO white dwarf to reach carbon ignition at the 
center ($\rho_\mathrm{c} = 2\times10^9~\mathrm{g~cm^{-3}}$), and 
electron-capture induced collapse ($\rho_\mathrm{c} = 
10\times10^{10}~\mathrm{g~cm^{-3}}$), respectively, according to 
\citet{Yoon05}.  A SN Ia explosion is expected in the hatched region. 
The filled circle denotes the initial model of the central remnant in 
Seq.~Sa.  The evolution of the central remnant in Seqs~Sa 8, Sa9 and 
Sa11 is shown by the thin dotted curves, as indicated. The thin solid 
curves denote possible evolutionary paths of the central remnant with 
strong magnetic torques that may enforce rigid rotation, with loss of 
angular momentum but without mass shedding (Case 'a'), and with both 
loss of angular momentum and mass shedding (Case 'b').  See the text 
for more details.  }\label{figmj} 
\end{center} 
\end{figure}

\begin{figure} 
\begin{center} 
\resizebox{0.8\hsize}{!}{\includegraphics{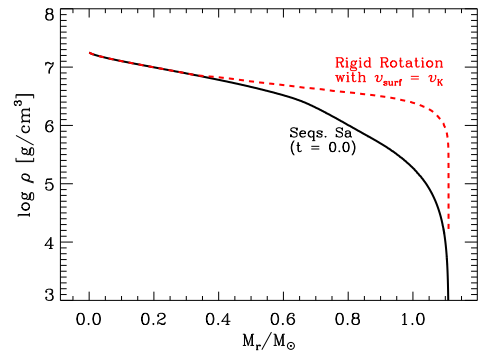}} 
\caption{The density profile in the initial model of the central remnant 
in Seqs~Sa (solid curve), and in a corresponding hot ($T_\mathrm{c} = 
10^8~\mathrm{K}$) white dwarf model that rotates rigidly at critical rotation 
at the surface (dashed curve).}\label{comp_rigid} 
\end{center} 
\end{figure}

The central remnant may be enforced to rotate rigidly on a short time 
scale in the presence of strong magnetic torques 
(cf. \citealt{Spruit02}).  The central remnant in both our SPH 
simulation and 1-D models has $J_\mathrm{tot} > 
10^{50}~\mathrm{erg~s}$, which is significantly higher than the 
maximum limit a rigidly rotating white dwarf can retain, as shown 
in Fig.~\ref{figmj}.  This means that if magnetic torques led to rigid 
rotation, a large amount of angular momentum should be transported 
into the Keplerian disc (Case \emph{a} in Fig.~\ref{figmj}), or mass 
shedding of super-critically spun-up layers should occur from the 
central remnant (Case \emph{b} in Fig.~\ref{figmj}).  In Case 
\emph{a}, the local density around the interface should increase by 
several factors by the time when the central remnant reaches rigid 
rotation as implied by Fig.~\ref{comp_rigid}.  Off-center carbon 
ignition might be inevitable in this case due to a resultant high 
effective accretion rate, if the time for angular momentum 
redistribution were shorter than the local cooling time due to 
neutrino losses.  In Case \emph{b}, on the other hand, the local 
density at the interface might not increase if mass shedding from the 
central remnant occurred at a sufficiently high rate.  Therefore, the 
role of magnetic fields in the merger evolution remains uncertain at 
the current stage and is a challenging subject for future work. 

The coalescence of more massive double CO white dwarf binaries is 
likely to result in a higher maximum temperature due to the enhanced 
role of gravity.  Consequently, given the important role of the 
maximum temperature in the merger remnant for its final fate, less 
massive binary CO white dwarfs may be favored for the production of 
SNe Ia from such a channel. 
 
We note that there are number of potentially important factors 
that have not been included in either the present study or 
previous simulations. These include the following points: 
\begin{enumerate} 
\item The previous and present simulations assumed that white dwarfs 
are cold prior to the merging process.  However, \citet{Iben98} point 
out that tidal interactions might heat up the white dwarfs as the orbit 
shrinks, which could weaken the gravitational potential of the 
primary. Furthermore, as the temperature of white dwarfs is a function of 
their age, younger progenitors should have more extended envelopes, which 
may result in a lower $T_\mathrm{p}$. 
 
\item A thin hydrogen/helium envelope must be present initially in 
both the primary and the secondary. As hydrogen or helium should 
ignite at a much lower temperature than carbon, the influence of the 
release of nuclear energy during the merger process may be even more 
important than shown in the existing SPH simulations, which is likely 
to lower $T_\mathrm{p}$. Furthermore, neutrino
losses, which were neglected in the present study, 
would also tend to reduce $T_\mathrm{p}$.

\item At a given total mass ($M_\mathrm{tot} = M_\mathrm{primary} + 
M_\mathrm{secondary}$), different mass ratios of the white dwarf 
components ($q\equiv M_\mathrm{secondary}/M_\mathrm{primary}$) must result 
in different merger structures. 
 
\item A lower $q$ at a given $M_\mathrm{tot}$ may not only lead to a 
stronger gravitational potential of the primary, but also to a lower 
mass accretion rate during the dynamical mass transfer 
\citep{Guerrero04}.  As the former and the latter will tend to 
increase and decrease $T_\mathrm{p}$, respectively, quantitative 
studies are necessary to predict how $T_\mathrm{p}$ will change 
with $q$. 
 
\end{enumerate} 
 
Finally, another important ingredient that needs to be considered is 
thermal diffusion during the dynamical evolution.  As shown above, the 
mass-accretion rate from the Keplerian disc onto the envelope of the 
central remnant is one of the most important factors that critically 
determine the final fate of double CO white dwarf mergers.  The accretion 
rates depend on the structure of the Keplerian disc at thermal 
equilibrium, which can be only understood by including thermal 
diffusion in future simulations. But here we emphasize again that 
the accretion rates from a centrifugally supported Keplerian disc should 
be significantly lower than those from a pressure-supported thick disc 
that was previously assumed, which opens the possibility for at least 
some double CO white dwarf mergers to produce SNe Ia.

\section*{Acknowledgments} 
 
We are grateful to Norbert Langer and Ken'ichi Nomoto for many useful suggestions and comments. 
SCY is supported by the VENI grant (639.041.406)  
of the Netherlands Organization for Scientific Research (NWO). 
The computations have been performed on the JUMP supercomputer at the H\"ochstleistungsrechenzentrum J\"ulich.

\end{document}